\documentclass[12pt, letterpaper]{article}
\pdfoutput=1
\usepackage{amsfonts}
\usepackage{amsmath}
\usepackage{amsthm}
\usepackage{bibentry}
\usepackage{color}
\usepackage{epsfig}
\usepackage{framed}
\usepackage{graphics}
\usepackage{hyperref}
\hypersetup{colorlinks = true, citecolor = blue, urlcolor = blue}
\usepackage[utf8]{inputenc}
\usepackage[left, modulo]{lineno}
\usepackage{lmodern}
\usepackage{multirow}
\usepackage{relsize}
\usepackage{rotating}
\usepackage{subfigure}
\usepackage{subfiles}
\usepackage{url}
\usepackage{verbatim}
\usepackage{wrapfig}
\usepackage{floatrow}
\newfloatcommand{capbtabbox}{table}[][\FBwidth]

\usepackage[round]{natbib}
\usepackage[noend]{algpseudocode}
\usepackage{algorithm}

\usepackage{graphicx,psfrag,epsf}
\usepackage{enumerate}

\makeatletter
\@addtoreset{equation}{section}   

\makeatother

\makeatletter
\@addtoreset{figure}{section}
\renewcommand{\thefigure}{\thesection.\@arabic\c@figure}
\makeatother

\makeatletter
\@addtoreset{table}{section}
\renewcommand\thetable{\thesection.\@arabic\c@table}
\makeatother



\usepackage{mycommands2}

\theoremstyle{definition}
\newtheorem{Algorithm}{Algorithm}
\newtheorem{Example}{Example}[section]

\begin{document}

\def\spacingset#1{\renewcommand{\baselinestretch}%
{#1}\small\normalsize} \spacingset{1}


  \title{\bf Fast and General Best Subset Selection using Data Depth and Resampling}
  \author{Subhabrata Majumdar\thanks{
    The author
acknowledges the University of Minnesota  Interdisciplinary Doctoral Fellowship
program. }\hspace{.2cm}\\
    University of Florida Informatics Institute\\
    and \\
  Snigdhansu Chatterjee\thanks{
    This research is partially  supported by
the National Science Foundation (NSF) 
under grants
 \# IIS-1029711 and 
\# DMS-1622483
and by the National Aeronautics and Space Administration (NASA).  }\hspace{.2cm}\\
    School of Statistics, University of Minnesota\\
    }
    \date{}
    \maketitle
    
\begin{abstract}
We present a technique using data depth functions and resampling to perform best subset variable selection for a wide range of statistical models. We do this by assigning a score, called an $e$-value, to a candidate model, and use a fast bootstrap method to approximate sample versions of these $e$-values. Under general conditions, $e$-values can separate statistical models that adequately explain properties of the data from those that do not. This results in a fast algorithm that fits only a single model and evaluates $p +1$ models, $p$ being the number of predictors under consideration, as opposed to the traditional requirement of fitting and evaluating $2^{p}$ models. We illustrate in simulation experiments that our proposed method typically performs better than an array of currently used methods for variable selection in linear models and fixed effect selection in linear mixed models. As a real data application, we use our procedure to elicit climatic drivers of Indian summer monsoon precipitation.
\end{abstract}

\noindent%
\vfill

\newpage

\section{Introduction}\label{sec:IntroSection}
\label{Section:Introduction}

Consider the linear model
\begin{align}\label{eqn:StartEqn}
Y = X \beta + \epsilon
\end{align}
where $Y \in \BR^n$ is the vector of responses, $X \in \BR^{n \times p}$ is the predictor matrix, $\beta \in \BR^p$ is the unknown vector of coefficients, and $\epsilon \in \BR^n$ is a vector of independent and identically distributed random errors. Many important applications of this model demand the detection of important predictors affecting the response and thus obtaining a parsimonious fit to the data in hand. Because of this, variable selection in linear models has been a well-studied topic, including classics \citep{ref:Akaike_AISM70203, ref:Schwarz_AoS78461,ref:Shao_JASA96655,Tibshirani96} and more recent advances \citep{BertsimasEtal16,LaiHannigLee15,Zou06}. The two major avenues variable selection has been explored through are sparse penalized regression and best subset selection. Sparse penalized regression has received abundant attention in the past two decades due to its relative computational simplicity and ability to tackle high-dimensional data. However, there are several inferential and algorithmic issues associated with such methods: for example the bias and confidence intervals of Lasso estimates \citep{ZhangZhang14} and use of convex relaxations to compute approximate local solutions of nonconvex penalized regression problems \citep{WangKimLi13,ZouLi08}. A detailed discussion of the caveats of penalized regression procedures can be found in \cite{BertsimasEtal16}.

On the other hand, the major challenge of performing best subset selection by checking multiple model fits is computational. In the setup of \eqref{eqn:StartEqn} there are $2^{p}$ potential models, and navigating through this model space is an NP-Hard problem \citep{Natarajan95}. Greedy methods like forward selection or backward deletion give no theoretical guarantee to select the correct model, and popular algorithms like leaps-and-bounds or branch-and-bounds are not computationally scalable to even moderate data dimensions ($n \simeq 100, p \simeq 50)$. A recent paper by \cite{BertsimasEtal16} proposed a tractable solution to the best subset problem using techniques from the optimization literature. It requires specifying the desired level of sparsity in the final model beforehand and works well in practice when the true model has a small number of non-zero coefficients. However, this is a largely unverifiable assumption, and there are more 	computational issues associated with it in high-dimensional situations, which we shall discuss later in the paper.



Although there has been some progress in best subset model selection for linear regression, it has received lesser attention for more complex models due to its perceived computational intractability and lack of intelligent search strategies. The fence methods \citep{ref:Jiangetal_AoS081669, NguyenJiang14} are popular in performing model selection in mixed effect models. Model selection criteria that expand the Akaike Information Criterion (AIC), Bayesian Information Criterion (BIC) and Mallow's $C_p$ to mixed models \citep{MezaLahiri05, VaidaBlanchard05} and robust regression \citep{RonchettiChapter} have also been proposed. However, they still depend on multiple model fits and suffer from the same scalability issues as linear regression best subset selection.

\paragraph{Our method}
In this paper, we propose a unified framework that is able to perform best subset selection in the coefficient vector $\beta$ for a range of statistical models much broader than \eqref{eqn:StartEqn} by traversing a very small proportion of the space of candidate models. Even though our proposal is not based on penalization, we are able to determine important variables in a model {\it by training a single model}: the one with all covariates under consideration. We do this by obtaining a score for a candidate model $\cM$ based on the amount of similarity between the \textit{sampling distribution} of the model coefficient estimate $\hat \beta_\cM$ and that of the least parsimonious estimate, say $\hat \beta$. We use data depths, which are functions that provide center-outward ranking for points in multivariate space with respect to a probability distribution \citep{tukey75,zuo03, ref:ZuoSerfling_AoS00461}, to compute this score- which we call the {\bf $e$-value} of the candidate model. 

Specifically, given some depth function $D(.,.)$ we define the $e$-value as
$$
e (\cM) = \BE D( \hat \beta_\cM, [ \hat \beta])
$$
i.e. the expected depth of the model estimates with respect to the sampling distribution of $\hat \beta$, which is denoted by $[\hat \beta]$. In this setup, for large enough $n$ the index set $\cS_{select} \subseteq \{ 1, \ldots, p \}$ obtained using Algorithm~\ref{alg:algoselect} elicits all non-zero covariates in the true parameter vector.
\begin{Algorithm}
\label{alg:algoselect}
(Best subset selection using $e$-values)

\noindent 1. Obtain $e$-value for the full model: $e(\cM_{full}) = \BE D ( \hat\bfbeta, [ \hat \bfbeta] )$.

\noindent 2. Set $\cS_{select} = \phi$.

\noindent 3. For $j$ in $1:p$

\hspace{1em} Replace $j^\text{th}$ index of $\hat\bfbeta$ by 0, name it $\hat \beta_{-j}$.

\hspace{1em} Obtain $e(\cM_{-j}) = \BE D ( \hat\bfbeta_{-j}, [ \hat \bfbeta] )$.

\hspace{1em} If $e(\cM_{-j}) < e(\cM_{full}))$

\hspace{2em} Set $ \cS_{select} \leftarrow \{ \cS_{select}, j \} $.
\end{Algorithm}
\noindent For a size-$n$ finite sample, we use resampling to generate multiple copies of $\hat \beta$ and subsequently estimate $e(\cM)$ for each model.

\begin{figure}[t]
\centering
\includegraphics[width=.4\textwidth]{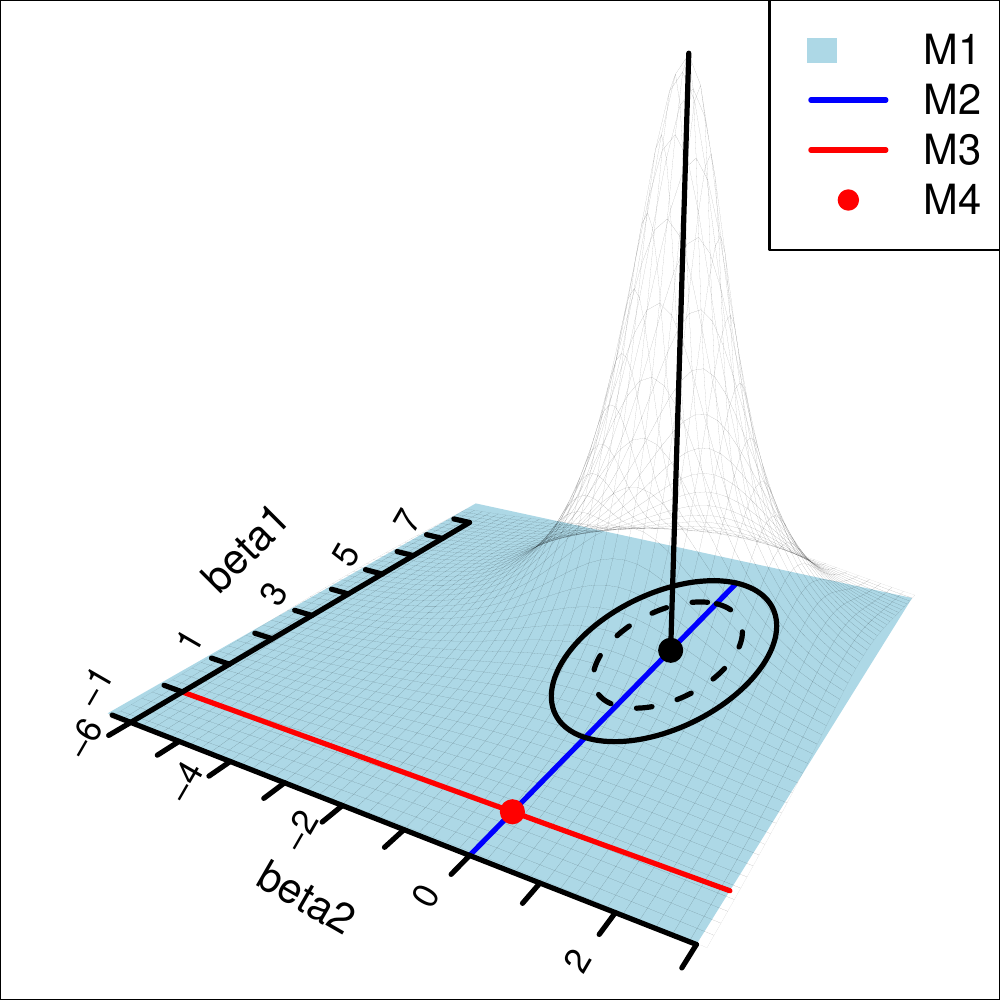}
\caption{Illustration of the $e$-value method. The depth surface is obtained using mahalanobis depth. The solid and dotted ellipses represent depth contours at some fixed level $\alpha >0$ for sample sizes $n_1$ and $n_2$, respectively, with $n_2 > n_1$.}
\label{fig:exampleFig}
\end{figure}

The example in Figure~\ref{fig:exampleFig} illustrates our motivation of using mean depth as a model selection criterion. Suppose $p=2$ and $\epsilon_i \sim \mathcal N (0, \sigma^2), \sigma>0, i = 1, \ldots, n$ in (\ref{eqn:StartEqn}), and consider the following choice of models:

\begin{center}
\begin{tabular}{ll}
$\cM_1: \ \ Y = X_1 \beta_1 + X_2 \beta_2 + \epsilon$; &$\ \ \Theta_1 = \BR^2$\\
$\cM_{2}: \ \ Y  =  X_1 \beta_1 + \epsilon$; &$\ \ \Theta_2 = \BR \times \{ 0 \}$ \\
$\cM_{3}: \ \ Y  =  X_2 \beta_2 + \epsilon$; &$\ \ \Theta_3 = \{ 0 \} \times \BR$\\
$\cM_{4}: \ \ Y  =  \epsilon $; &$\ \ \Theta_4 = (0,0)^T$
\end{tabular}
\end{center}
\noindent where $\Theta_i$ denotes the domain of parameter values corresponding to model $\cM_i; i = 1, ..., 4$. Now assume that the true value of the coefficient vector is $\beta_0 = (5,0)^T$, $a^T$ denoting the transpose of a vector or matrix-valued $a$. The full model estimate has sampling distribution $\mathcal N(\beta_0, \sigma^2 (X^T X)^{-1})$. As $n$ grows, this distribution concentrates around $\beta_0$, and the depths at points along the (red) line $\beta_2=0$, as well as at the (red) point (0,0) become smaller and smaller. Hence, the expected depths go to 0 for $\cM_3$ and $\cM_4$. However, owing to affine invariance of depth functions, the range of depth values for points in $\cM_1$ (blue line) and $\cM_2$ (blue surface) stays the same. Moreover, depth contours coincide with density contours for elliptical distributions \citep{LiuSingh93}, thus $e(\cM_1)$ and $e(\cM_2)$ remain positive and do not converge to 0 as $n \rightarrow \infty$.

For the purpose of covariate selection, the space of all possible models can be divided into two classes: models with non-zero support that contains the non-zero support of the true coefficient vector, and models that do not satisfy this. Because of the above reasoning, $e$-values for models in the first class become clearly separated from $e$-values of models in the second class. We further leverage the properties of depth functions to prove that within the first class of `good' models, a rank ordering of $e$-values of the models exists based on the amount of parsimony in them: after which Algorithm \ref{alg:algoselect} becomes immediate.


\paragraph{Structure of paper} The paper is organized as follows. Section~\ref{sec:PrelimSection} provides background for the development of our framework. We define $e$-values formally in Section~\ref{sec:evalueSection}, present details about their estimation through resampling and their implementation through our fast algorithm. Theoretical results are presented in Section~\ref{sec:Theory}. In Section~\ref{Section:Simulation} we present illustrative examples on how our fast model selection algorithm is implemented, and its performance relative to state-of-art methods in both classical $n>p$ and high-dimensional situations.  This is followed in Section~\ref{Section:IndianMonsoon} with a study on the possible determinants of precipitation in the Indian sub-continent during the summer monsoon months. In Section~\ref{Section:Conclusion} we discuss the potentials of our proposed method, directions of future research, as well as some caveats and concluding comments. Proofs of all theoretical results are given in the appendix.

\paragraph{Notations} We denote the $(i,j)^\text{th}$ entry of a matrix $M \in \BR^{A \times B}$ by $(M)_{ij}$. The notations $I_A$ and $0_A$ indicate the identity matrix of dimension $A \times A$ and the vector of zeros of dimension $A$, respectively. For a vector $v = (v_1, \ldots, v_A)^T \in \BR^A$ we denote its euclidean norm by $\| v \| = \sqrt{\sum_{i=1}^A v_i^2}$, and $l_0$-norm by $\| v \|_0 = \sum_{i=1}^A \BI (v_i \neq 0)$: $\BI(.)$ being the indicator function. We denote the non-zero support of $v$ by $\supp(v)$, i.e. $\supp(v) = \{ i: v_i \neq 0 \}$. For random variables $R, R_n, n = 1, 2, \ldots$, the notation $R_n \leadsto R$ means convergence in distribution, and $R_n \stackrel{P}{\rightarrow} R$ or $ \text{plim}_{n \raro \infty} R_n = R$ means convergence in probability. Expectations are denoted by $\BE(\cdot)$, and the variance of a scalar-valued random variable or the covariance matrix of a vector-valued random variable is denoted by $\BV(\cdot)$.
\section{Preliminaries}
\label{sec:PrelimSection}
We introduce the theoretical framework necessary for our methodology in this section. This includes specifications for the following three things: (1) an appropriate method of estimation, (2) categorization of candidate models, and (3) the choice of a depth function.

Suppose at stage $n$ there is a triangular array of functions
\baq\label{eq:PsiEqn}
\{ \Psi_{ni} (\theta_n, Z_{ni}): 1 \leq i \leq k_n, n \geq 1 \}
\eaq
where $\cZ_n = \{ Z_{n1}, \ldots, Z_{n k_n} \}$ is an observable array of random variables, and $\theta_n \in \Theta_n \subseteq \BR^{p}$. Functions like $\Psi_{n i} (\cdot)$ have been referred to as {\textit{energy functions}} in  optimization and related literature, and have also been called {\textit{contrast functions}} \citep{ref:BoseC_JSPI03225, ref:MichelPfanzagl_ZWVG7173, ref:Pfanzagl_Metrika69249}. This framework naturally includes different kinds of regressions, mixed effect models, times series and other traditional topics where covariate or model selection has been historically studied. Note that we use a triangular array-based framework here, where the energy functions and the true parameter may depend on $n$ for precise theoretical study, and to capture the scientific reality that data, models and goals of research evolve over time.

We assume that there is a true unknown vector of parameters $\theta_{0 n}$, which is the unique minimizer of
\begin{align}
\Psi_n (\theta_n) = \BE \sum_{i=1}^{k_n} \Psi_{n i} (\theta_n, Z_{n i})
\label{eq:Psistarn}
\end{align}
Also denote the common non-zero support of all estimable parameters as $\cS_{* n} = \cup_{ \theta_n \in \Theta_n} \supp(\theta_n)$. In this general setup, we associate a candidate model $\cM_n$ with two quantities:

\begin{enumerate}
\item[(a)] The set of indices $\cS_n \subseteq \cS_{* n} $ where the parameter values are unknown and estimated from the data, and
\item[(b)] an ordered vector of known constants $C_{n} = (C_{n j}: j \notin \cS_n)$ for  parameters not indexed by $\cS_n$.
\end{enumerate}
%
%

\noindent The generic parameter vector corresponding to this model, denoted by $\theta_{m n} \in \Theta_{m n} \subseteq \Theta_{n} := \prod_j \Theta_{n j}$, will thus have the structure
\baq 
 \vectheta_{m n j} = \left\{ \begin{array}{ll}
 \text{ unknown} \ \theta_{ m n j} \in \matTheta_{n j} & \text{ for } 
 			j \in \cS_n, \\
 \text{ known} \  C_{n j} \in \matTheta_{n j} & \text{ for } j \notin \cS_n.
\end{array}
\right.
\label{eq:vectheta_mn}
\eaq
So that in each model a part of the parameter vector consists of known constants, and the other part is unknown.

\subsection{Method of estimation}
The estimator $\hat{\theta}_{* n}$ of $\theta_{0 n}$ is obtained as a minimizer of the sample analog of (\ref{eq:Psistarn}). Thus
\begin{align}
\hat{\theta}_{* n} = 
\argmin_{\theta_n} \hat{\Psi}_{n} (\theta_n) =
\argmin_{\theta_n} \sum_{i = 1}^{k_{n}} \Psi_{n i}  \bigl( \theta_n, Y_{ni} \bigr)
\label{eq:Psistarnhat}
\end{align}
We assume an elliptical asymptotic distribution for $\hat \theta_{* n}$. Following \cite{FangEtalBook}, the density function of an elliptically distributed random variable takes the form:
$$ h(x; \mu, \Sigma) = |\Sigma|^{-1/2} g ((x - \mu)^T \Sigma^{-1} (x - \mu)) $$
where $\mu \in \BR^p$, $\Sigma \in \BR^{p \times p}$ is positive semi-definite, and $g$ is a non-negative scalar-valued density function that is continuous and strictly increasing, and is called the \textit{density generator} of the elliptical distribution. We denote such an elliptical distribution by $\cE (\mu, \Sigma, g)$. For the asymptotic parameter distribution we also assume the following conditions:

\begin{enumerate}
\item[(A1)]
There exists a sequence of positive reals $a_n \uparrow \infty$ such that $a_{n} (\hat \bftheta_{* n} - \bftheta_{0 n} ) \leadsto \cE ( 0_p, \bfV, g)$, for some positive-definite matrix $\bfV \in \BR^{p \times p}$ and density generator function $g$;

\item[(A2)]
For almost every data sequence $\cZ_n$, There exists a sequence of positive definite matrices $\bfV_n \in \BR^{p \times p}$ such that  $\text{plim}_{n \raro \infty} \bfV_n = \bfV$.
\end{enumerate}

Note that under standard regularity conditions the limiting distribution of a suitably centered and scaled parameter estimator will be Gaussian, in which case assumptions (A1)-(A2) are naturally satisfied.

The estimator in \eqref{eq:Psistarnhat} corresponds to the {\it full model} $\cM_{* n}$, i.e. the model where all indices in $\cS_{* n}$ are estimated. For any other model $\cM_n$,
%
%
we simply augment entries of $\hat \theta_{* n}$ at indices in $\cS_n$ with elements of $C_n$ elsewhere to obtain a model-specific coefficient estimate:
\baq 
 \hat{\vectheta}_{m n j} = \left\{ \begin{array}{ll}
 \hat{\theta}_{ * n j} & \text{ for } j \in \cS_n, \\
 C_{n j} & \text{ for } j \notin \cS_n.
\end{array}
\right.
\label{eq:thetahat_mn}
\eaq
The logic behind this plug-in estimate is simple: for a candidate model $\cM_{n}$, a joint distribution of its 
estimated parameters, i.e. $[\hat{\vectheta}_{s n}]$, can actually be obtained from $[ \hat{\vectheta}_{* n}] $ by marginalizing at indices $\cS_n$. 


\subsection{Characterization of models}
Given the formulation of models in \eqref{eq:vectheta_mn}, any model can be classified into two categories: {\it adequate} and {\it inadequate} models.

\begin{Definition}
A model $\cM_{n}$ is called  
{\textit{adequate}} if 
\baq
\lim_{n \raro \infty}
\sum_{j \notin \cS_n} | C_{n j}  - {\theta}_{0 nj} | = 0
\label{eq:adequateModel}
\eaq
%
A model that is not adequate, will be called an {\textit{inadequate}} model.
\end{Definition}
We also consider a stronger version of adequacy below.
\begin{Definition}
The model $\cM_{n}$ is called {\textit{strictly adequate}} if, for all $n$ and $j \notin \cS_n$, $C_{n j}  = {\theta}_{0 n j}$.
\end{Definition}

By definition the full model is always adequate, as is the model corresponding to the singleton set containing the true parameter, i.e. $\{ \theta_{0 n} \}$ (because $\supp(\theta_{0 n}) \subseteq \cS_{* n}$). Thus the set of adequate models is non-empty by construction.

In classical variable selection where all elements of the constant terms $C_n$ are set to 0 for all models, as in linear regression where a subset of covariates $X_{s}$ is used in fitting the expression $Y = X_{s} \beta_{s} + \epsilon_{s}$, the above notion of model adequacy captures standard notions of model `correctness'. Specifically, 
\eqref{eq:adequateModel} reduces to $\lim_{n \raro \infty} \BE Y - X_{s} \beta_{s} = 0$. This is weaker than the traditional condition $\BE Y - X_s \beta_s = 0$ about model `correctness' implicit in many studies, which is captured in our condition for a model being {\it strictly adequate}. In addition, our definition allows asymptotically similar models to be clubbed together, and has the flexibility to accommodate local asymptotic characteristics. For instance, consider the data-generating model $Y_{ni} = X_{1i} \beta_{01} + X_{2 i} \delta_n + \epsilon_i$ for some $\beta_{01} \in \BR, \delta_n = o (1), \epsilon_{i} \iid N(0, \sigma_0^{2})$ with $\sigma_0 > 0$ and $i = 1, \ldots, k_n$. Evidently, here $\theta_{0 n} = (\beta_{0 1}, \delta_n, \sigma_0)^T$. With our formulation, the (constant) sequence of models $\cM_n$ with $\bfTheta_{m n} = \{ (\beta_1, 0, \sigma^{2})^T: \beta_1 \in \BR, \sigma > 0 \}$ will be adequate. Such situations can arise from prior choices in Bayesian variable selection techniques, for example see \cite{NarisettyHe14,RockovaGeorge16}.

Besides model adequacy, a notion that will be important in our framrwork is the one of \textit{nested models}.
\begin{Definition}
We consider a model $\cM_{1 n}$ to be nested in $\cM_{2 n}$, notationally $\cM_{1 n} \prec \cM_{2 n}$, if 
$\cS_{1 n} \subset \cS_{2 n}$ and $C_{2 n}$ is a subvector of $C_{1 n}$.
\end{Definition}
\noindent If a model $\cM_{n}^{ad}$ is adequate, then any model $\cM_n$ such that $\cM_n^{ad} \prec \cM_n$ is also adequate. In the context of covariate selection in the linear model of \eqref{eqn:StartEqn}, this obtains a linear ordering, with the most parsimonious adequate model having $\cS_n = \supp (\beta_0)$ and $C_n = 0_{p - | \cS |}$ nested in all other adequate models. However all models nested {\it within} it shall be inadequate. 

\subsection{Depth functions}
Let $\cG$ denote the set of probability measures on $\BR^{p}$, and let $D : \BR^{p} \times \cG \raro [0, \infty)$ be a data depth function \citep{ref:ZuoSerfling_AoS00461}. We consider data depth functions satisfying the following properties:
\begin{enumerate}
\item[(B1)]
The map $D$ is invariant to affine transformations, i.e. for any non-singular matrix $\bfA \in \BR^{p \times p}$, and $b \in \BR^p$ and random variable $Y$ having distribution $\BG \in \cG$, 
$$
D (x, \BG) = D ( \bfA x + b, [ \bfA Y + b ])
$$
\item[(B2)]
The map $D(., \BG)$ is Lipschitz continuous, where $\BG \in \cG$ is a fixed, non-degenerate probability distribution. That is, there exists $\delta > 0$ and $\alpha \in (0, 1)$, possibly depending on $\BG$ such that whenever  $| x - y | < \delta$, we have 
$$ 
| D (x, \BG) - D (y, \BG) | < | x - y|^{\alpha} 
$$

\item[(B3)]
Assume that $Y_{n} \in \BR^{p}$ is a sequence of random variables converging in distribution to some $\BY \in \cG$. Then $D (y, [Y_n])$  converges uniformly to $D (y, \BY) $. In particular, if $Y \sim \BY$, then 
$$
\lim_{n \raro \infty} \BE D (Y_n, [Y_n]) = \BE D (Y, \BY)
$$
exists and is finite.

\item[(B4)]
For any $\BG \in \cG$, $\lim_{\| \bfx \| \raro \infty} D( \bfx, \BG) = 0$.

\item[(B5)]
For any  $\BG \in \cG$ with a point of symmetry $\bfmu (\BG) \in \BR^{p}$, we have for any $ t \in (0, 1)$ and any $\bfx \in  \BR^{p}$ 
$$
D ( \bfx, \BG) < D (\bfmu (\BG) + t (\bfx - \bfmu (\BG)), \BG)
< D (\bfmu (\BG), \BG)
= \sup_{\bfx \in  \BR^{ p}} D ( \bfx, \BG) < \infty
$$
That is, the depth function takes a maximum value at $\bfmu (\BG)$, and is strictly decreasing along any ray connecting $\bfmu (\BG)$ to any point $\bfx \in  \BR^{ p}$.
\end{enumerate}
Conditions (B1), (B4) and (B5) are integral to the formal definition of data depth \citep{ref:ZuoSerfling_AoS00461}, while (B2) and (B3) implicitly arise for several depth functions \citep{MoslerChapter13}. For most of the theoretical analysis in this paper we require only a subset of these properties. However, we shall continue to use data depths throughout for simplicity.
\section{The e-values methodology}
\label{sec:evalueSection}
We now associate with each model $\cM_{n}$ a functional of the depth of the model estimate with respect to the full model sampling distribution. We call this the \textbf{e-value} of the model. An example of $e$-value is the mean depth: 
\begin{equation}
e_n (\cM_n) = \BE  D (  \hat \theta_{ m n}, [\hat \theta_{*n}] )
\label{eq:Dsn}
\end{equation}
which we concentrate on for the rest of the paper. However, any other functional of $D (  \hat \theta_{ m n}, [\hat \theta_{*n}] )$ may also be used here, and a large proportion of our theoretical discussion is applicable to any smooth functional of the distribution of $D (  \hat \theta_{ m n}, [\hat \theta_{*n}] )$. Furthermore, the distribution of $D (  \hat \theta_{ m n}, [\hat \theta_{*n}] )$ is itself informative, and has an important role to play in the study of uniform convergence. We defer all this discussion and analysis to future research.



\subsection{Resampling approximation of $e$-values}
\label{sec:BootSection}
There are two random quantities involved in \eqref{eq:Dsn}, namely $\hat \theta_{m n}$ and $\hat \theta_{* n}$. Typically, the distribution of either of them is not known, and have to be elicited from the data. We use resampling for this purpose. Because of the plugin method in \eqref{eq:thetahat_mn}, only $[ \hat \theta_{* n}]$ needs to be approximated, and we propose using Generalized Bootstrap \citep{ref:CBose_AoS05414} for this. We take an exchangeable array of  non-negative random variables, independent of the data, as resampling weights: $\cW_{r n} = ( \BW_{r n 1}, \ldots,  \BW_{r n k_{n}})^{T} \in \BR^{k_{n}}$, and obtain the resampling estimator $\hat{\vectheta}_{r * n}$ as the minimizer of 
\baq
\hat{\Psi}_{r n} (\theta_n) 
= \sum_{i = 1}^{k_{n}} \BW_{r n i}\Psi_{n i}  \bigl( \theta_n, Z_{n i} \bigr).
\label{eq:Psisnhat_R}
\eaq
We assume the following conditions on the resampling weights as $n \raro \infty$:
\baq
\BE \BW_{r n 1}  & = 1, 
\label{eq:ResamplingWeightMean}\\
\BV \BW_{r n 1}  & = \tau_{n}^{2} \uparrow \infty, 
\label{eq:ResamplingWeightVariance}\\
\tau_{n}^{2} 
& = o ( a_{n}^{2}), 
\label{eq:TauSqCondition}\\
\BE W_{r n 1} W_{r n 2} & = O (k_{n}^{-1}), 
\label{eq:c11} \\
\BE W_{r n 1}^{2} W_{r n 2}^{2} & \raro 1, 
\label{eq:c22} \\
\BE W_{r n 1}^{4} & < \infty. 
\label{eq:c4}
\eaq
Several resampling schemes can be described in the above format and are discussed in \cite{ref:CBose_AoS05414}. Here we provide two examples.

\begin{Example}[The $m$-out-of-$n$ or {\it moon}-bootstrap]
\label{Example:BootExample1}

In our framework, the \textit{moon}-bootstrap is identified with $\cW_{r n}$ having a Multinomial distribution with parameters $m$ and probabilities $k_{n}^{-1}(1, \ldots, 1) \in \BR^{k_{n}}$, by a factor of $k_{n}/m$.  Thus we have $\BE \BW_{r n i} = (m^{-1} k_{n}) (m/k_{n}) = 1$, and $\BV \BW_{r n i} = \tau_{n}^{2} = (m^{-1} k_{n})^{2} (m k_{n}^{-1} (1 - k_{n}^{-1}) = O( m^{-1} k_{n})$. In typical applications of the \textit{moon}-bootstrap, we require that $m \raro \infty$ and $m/k_{n} \raro 0$ as $n \raro \infty$. Thus we have $\tau_{n}^{2}  \raro \infty$ as $n \raro \infty$, that is, the scale factor of the resampling weights $\BW_{r n i}$ tending to infinity with $n$.
\qed
\end{Example}

\begin{Example}[The scale-enhanced Bayesian bootstrap]
\label{Example:BootExample2}
A version of Bayesian bootstrap may be constructed by choosing $\BW_{r n i}$ to be independent and identically distributed Gamma random variables, with mean 1 and variance $\tau_{n}^{2}  \raro \infty$ as $n \raro \infty$. The functionality of this resampling scheme and Bayesian interpretation remain similar to the standard Bayesian bootstrap, however some convenient properties like conjugacy are lost.
\qed
\end{Example}
%
%
%
%
%
%
%

Under fairly weak regularity conditions on the first two derivatives of $\Psi_{ni}(\cdot)$, \cite{ref:CBose_AoS05414} proved that $(a_n/\tau_n) (\hat \theta_{r * n} - \hat \theta_{* n})$ converges to the same weak limit in probability as $a_n (\hat \theta_{* n} - \theta_{0 n})$. To estimate the $e$-value of any model we obtain two independent copies of the solution to \eqref{eq:Psisnhat_R}, which we index by $r$ and $r_{1}$. We use the first set of resamples to obtain the distribution $[\hat{\vectheta}_{r * n}]$ conditional on the data that approximates $[ \hat \theta_{* n} ]$, and the second set of resamples to obtain $\hat{\vectheta}_{r_{1} m n}$ corresponding to the model $\cM_{n}$:
\begin{align}\label{eqn:bootEstEqn}
 \hat{\theta}_{r_{1} m n j} = \left\{ \begin{array}{ll}
 	 \hat{\theta}_{r_1 * n j} & \text{ for } 
 			j \in \cS_n; \\
 	 C_{n j} & \text{ for } j \notin \cS_n
\end{array}
\right.
\end{align}
Consequently, the resampling estimate of a model $e$-value is defined as 
\baq 
e_{r n} (\cM_{n}) = \BE_{r_{1}} D \bigl(  \hat{\vectheta}_{r_{1} m n}, 
[\hat{\vectheta}_{r * n}] \bigr), 
\label{eq:BootEValue}
\eaq
where $\BE_{r_{1}}$ is expectation, conditional on the data, of the resampling indexed by $r_{1}$.

The actual implementation of the resampling procedure is carried out by generating independent copies $\cW_{1 n}, \ldots, \cW_{R n} $ for some sufficiently large integer $R$. We do not need to solve \eqref{eq:Psisnhat_R} explicitly to calculate $\hat \theta_{r * n}$. Rather, one can use already computed model quantities in the following approximation to get them:
\begin{align}\label{eqn:BootEqn}
\hat\theta_{r * n} & = \hat\bftheta_{* n} - \frac{ \tau_n}{a_n} \left[ \sum_{i=1}^n W_i \Psi_{n i}'' (\hat \bftheta_{* n}, Z_{n i}) \right]^{-1}
\sum_{i=1}^n W_i \Psi_{n i}' (\hat \bftheta_{* n}, Z_{n i}) + \bfR_{rn};\\
& \BE_r \| \bfR_{r n} \|^2 = o_P(1)\notag 
\end{align}
This is immediate from Theorem~3.5 in \cite{ref:CBose_AoS05414}. Consequently only Monte Carlo sampling is required to obtain the resamples, resulting in significant computational speed gains.

\subsection{Fast algorithm for best subset selection}
\label{subsec:Algo}
For best subset selection we consider the class of candidate models
\baq
\mathbb M_{0} = \{ \cM: C_j=0 \quad \forall \quad j \notin \cS_* \}
\label{eqn:M0definition}
\eaq
dropping ${}_n$ in the subscripts of $\cM_n, C_n, \cS_n, \theta_n$ (and their variants). Note that all adequate models are strictly adequate now, and their non-zero support contains $\supp(\theta_0)$. In this setup our fast selection algorithm consists of only three stages:

\begin{enumerate}
\item[(a)]
Fit the full model and estimate its $e$-value,
\item[(b)]
Replace each covariate by 0 and compute  $e$-value of all such reduced models, and
\item[(c)]
Collect covariates dropping which causes the $e$-value to go down.
\end{enumerate}

To compute the full model, we need to determine the estimable set of predictor indices $\cS_*$. In traditional $n > p$ situations, the choice is simple: $\cS_* = \{ 1, \ldots, p\}$. In high-dimnensional cases when $p > n$, we need to ensure that $p' = | \cS_* | < n$, so that $\hat \theta_{*}$ (properly centered and scaled) has a unique asymptotic distribution, i.e. satisfies condition (A1). We use fast screening methods available in the literature to achieve this, as implemented previously by \cite{LaiHannigLee15}. For high-dimensional linear regression we use the Sure Independent Screening (SIS) procedure \citep{FanLv08}, while for more complex cases like mixed models, single index models etc., we propose using the model-free screening method of \cite{ZhuEtal11}.

After obtaining $\cS_*$ and the full model estimate, for each of the $p'+1$ models under consideration: the full model and all dropped-predictor models, we generate two sets of resampling weights, use them in \eqref{eqn:BootEqn} to obtain bootstrapped coefficient estimates, and finally calculate finite sample approximations of the bootstrap $e$-values in \eqref{eq:BootEValue}. This gives us all the components for a sample version of the population level algorithm \ref{alg:algoselect}. We present this as Algorithm \ref{alg:algoselectboot}.

\begin{Algorithm}\label{alg:algoselectboot}
(Best subset selection using $e$-values and bootstrap)

\noindent 1. Fix resampling standard deviation $\tau_n$.

\noindent 2. Obtain bootstrap samples: $\cT = \{ \hat\theta_{1 *}, ..., \hat\theta_{R *} \} $, and $\cT_1 = \{ \hat\theta_{1 *}, ..., \hat\theta_{R_1 *} \} $.

\noindent 3. Calculate $\hat e_{r n} (\cM_*) = \frac{1}{R_1} \sum_{r_1=1}^{R_1} D ( \hat\theta_{r_1 *}, [\cT_1] )$.

\noindent 4. Set $\hat \cS_0 = \phi$.

\noindent 5. For $j$ in $1:p$

\hspace{1em} For $r_1$ in $1:R_1$

\hspace{2em} Replace $j^\text{th}$ index of $\hat\theta_{* r_1}$ by 0 to get $\hat \theta_{r_1, -j}$.

\hspace{1em} Calculate $\hat e_{r n} (\cM_{-j}) = \frac{1}{R_1} \sum_{r_1=1}^{R_1} D ( \hat\theta_{r_1, -j}, [\cT_1] )$.

\hspace{1em} If $\hat e_{r n} (\cM_{-j}) < \hat e_{r n} (\cM_*)$

\hspace{2em} Set $ \hat \cS_0 \leftarrow \{ \hat \cS_0, j \} $.
\end{Algorithm}

\paragraph{Remark} Choice of the bootstrap standard deviation $\tau_n$ plays a central role in the performance of Algorithm \ref{alg:algoselectboot}. Its intermediate rate of divergence, i.e. $\tau_n \rightarrow \infty, \tau_n/a_n \rightarrow 0$, is a necessary and sufficient condition for the consistency of the generalized bootstrap procedure (see \cite{ref:CBose_AoS05414} for details), as well as that of the bootstrap approximation of population $e$-values (Theorem~\ref{Theorem:ModelScore}). Variable selection literature has several examples of such quantities in action that result in model selection consistency: for example the penalty term of BIC, or the tuning parameters of adaptive lasso \citep{Zou06} and Dantzig selector \citep{CandesTao07}. We demonstrate the performance of Algorithm \ref{alg:algoselectboot} with several choices of $\tau_n$ in our simulations, and use validation on a test set to select the optimal $\tau_n$ in our data example.
\section{Theoretical results}
\label{sec:Theory}

We investigate theoretical properties of $e$-values in this section. Our first results are in the most general setup of adequate and inadequate models. We then focus on strictly adequate models, and finally on $\mathbb M_0$, as a consequence deriving our fast algorithm.

\subsection{Separation of adequate models}
Our first result on the model elicitation process separates inadequate models from adequate models at the population level.
\begin{Theorem}\label{Theorem:ThmRightWrong}
Assume the conditions (A1)-(A2) on $\hat \theta_{* n}$, and that $\BE \| \hat{\theta}_{* n} \|^{4} < \infty$. 
For any data-depth function satisfying (B1)-(B4), the following hold as $n \rightarrow \infty$:
\begin{enumerate}
\item For the full model $\cM_{* n}$, $e_n (\cM_{* n}) \rightarrow e_0 < \infty$, where $e_0 = \BE D(Y, [Y])$ with $Y \sim \cE(0_p, V, G)$,

\item For an adequate model $\cM_n$, 
$\left| e_n \left( \cM_n \right) - e_n \left( \cM_{* n} \right) \right| \raro 0$,

\item For inadequate model $\cM_n$, $e_n (\cM_n) \raro 0$.
\end{enumerate}
\end{Theorem}
This ensures that for large enough $n$, it is possible to find some threshold $\epsilon_n \leq e_n(\cM_{* n})$ such that all inadequate models have $e$-values less than the threshold, while $e$-values for all adequate models fall above it.

For generalized bootstrap approximations of population $e$-values through \eqref{eq:BootEValue}, we have the following equivalent result:
\begin{Theorem}
\label{Theorem:ModelScore}
Assume conditions of Theorem~\ref{Theorem:ThmRightWrong}, and $\BE \| \hat{\theta}_{* n} \|^{8} < \infty$. Additionally, assume that the resampling weights $\BW_{r n i}$ are exchangeable random variables satisfying the conditions \eqref{eq:ResamplingWeightMean}-\eqref{eq:c4}. Then, conditional on the data, $e_{r n} (\cM_{n})$ converges to zero in probability if $\cM_{n}$ is a sequence of inadequate models, and to $e_0$ if it is a sequence of adequate models.
\end{Theorem}

\paragraph{Remark} An examination of the proofs of the above two results, which follow similar lines, reveals that they do not use all properties of data depths. Specifically, only location and scale invariance is needed in place of the much stronger affine invariance in (B1), and condition (B5) is not needed at all. We also do not use condition (A2) for $\hat \theta_{* n}$. Consequently, theorems \ref{Theorem:ThmRightWrong} and \ref{Theorem:ModelScore} continue to hold for depth-like functions in a diverging-$p$ setup under relaxed versions of conditions (B1)-(B5) and some technical conditions on the energy functions, and possibly have broader implications than variable selection only. We plan to explore this in detail in future work.

\subsection{Derivation of algorithm}
\label{subsec:FastMS}
Our next result is concerned with strictly adequate models, and forms the foundation for the fast model 
selection algorithm.

\begin{Theorem}\label{Theorem:ThmRightNested}
Assume conditions (A1)-(A2), (B1)-(B5). Then, for a finite sequence of strictly adequate models $\cM_{1 n} \prec  \ldots \prec \cM_{k n}$ and any finite collection of inadequate models $\cM_{k +1, n}, \ldots, \cM_{K n}$, we have 
\ban 
 e_n(\cM_{1 n} ) > \ldots > e_n (\cM_{k n})  > \max_{j \in \{ k + 1, \ldots K \}} e_n (\cM_{j n})
\ean
for large enough $n$.
\end{Theorem}

We set the constant terms $C_n$ to 0 now, and drop the subscript ${}_n$ from $\cM_n, \cS_n, C_n, \theta_n$ and their variants. Recall that we have not yet used the actual data generating process as a candidate model. Let us define the \textit{data generating model} as $\cM_0 \prec \cM_*$ with associated estimable index set $\cS_0 = \supp (\theta_0)$ and $C_0 = 0_{p - | \cS_0|}$ for $j \notin \cS_0$. Then we have the following result.
\begin{Corollary}\label{Corollary:ZeroModelCorollary}
Assume the conditions of Theorem~\ref{Theorem:ThmRightNested}. Consider the collection of candidate models $\mathbb M_{0} = \{ \cM: C_j=0 \quad \forall \quad j \notin \cS_* \}$. Then for large enough $n$,
\begin{align}
\cM_{0} = \arg \max_{ \cM \in \mathbb M_{0}} \left[ e_{n} ( \cM ) \right]
\end{align}
\end{Corollary}

Thus, within the restricted set of models $\mathbb M_0$, the $e$-value maximizes at the true model. However there are still $2^p$ possible models. This is where the true advantage of using $e$-values comes through. It hinges on the following observation.

\begin{Corollary}\label{Corollary:AlgoCorollary}
Assume the conditions of Corollary~\ref{Corollary:ZeroModelCorollary}. Consider the models $\cM_{ -j} \in \mathbb M_0$ with  $\cS_{-j} = \{ 1, \ldots ,p \} \setminus \{ j \}$ for $j = 1, \ldots ,p$. Then covariate $j$ is a necessary component of $\cM_0$, i.e. $\cM_{-j}$ is an inadequate model, if and only if
\begin{align}
e_{n} (\cM_{-j}) < e_{n} (\cM_*) 
\end{align}
for sufficiently large $n$.
\end{Corollary}

Dropping an essential predictor from the full model makes the model inadequate, which has very small $e$-value for large enough $n$ (Theorem~\ref{Theorem:ThmRightWrong}), whereas dropping a non-essential predictor increases the $e$-value owing to Theorem~\ref{Theorem:ThmRightNested}. Thus, {\it simply collecting those predictors that cause decrease in the $e$-value on dropping them from the model suffices for variable selection}.

\subsection{Model selection consistency}
Consistency of bootstrap analogues of all model $e$-values follows from Theorem~\ref{Theorem:ModelScore}. This, in conjunction with the results from Section~\ref{subsec:FastMS} establishes model selection consistency of the finite sample bootstrap procedure.
\begin{Corollary}\label{corollary:BootConsistency}
Consider two sets of bootstrap estimates of $\hat \theta_*$: $\cT = \{ \hat \theta_{r *}: r = 1, \ldots, R \}$ and $\cT_1 = \{ \hat \theta_{r_1 *}: r_1 = 1, \ldots, R_1 \}$. Obtain sample $e$-value estimates as
\baq\label{eqn:BootEvalueMC}
\hat e_{r n} (\cM) & = \frac{1}{R_1} \sum_{r_1=1}^{R_1} D( \hat \theta_{r_1 m}, [ \cT ])
\eaq
where $[\cT]$ is the empirical distribution of the corresponding bootstrap samples, and $\hat \theta_{r_1 m}$ are obtained using \eqref{eq:BootEValue}. Consider the set of predictors $\hat \cS_0 = \{ \hat e_{r n} (\cM_{-j} ) < \hat e_{r n} (\cM_*) \}$. Then as $n, R, R_1 \rightarrow \infty$,

\baq\label{eqn:Consistency}
P_2( \hat \cS_0 = \cS_0 ) \rightarrow 1
\eaq
where $P_2$ is the probability conditional on the data and bootstrap samples.
\end{Corollary}
Equation~\eqref{eqn:Consistency} is contingent on the fact that the the data generating model $\cM_0$ is indeed a member of $\BM_0$, i.e. $\cS_0 \subseteq \cS_*$. This is ensured trivially in classical $n > p $ situations. If $p$ is larger than $n$, $\BM_0$ is the set of all possible models on the predictor set selected by an appropriate predictor screening procedure.

In high-dimensional linear models, we use the SIS procedure for this purpose. Given that the estimable index set $\cS_*$ is selected using SIS, \cite{FanLv08} proved under fairly weak conditions that
\baq\label{eqn:SISConsistency}
P( \cM_0 \in \BM_0) \geq 1 - O \left( \frac{\exp[ -C n^{1-2 \kappa} ]}{\log n} \right)
\eaq
where $C>0$ and $\kappa$ are constants that depend on the minimum signal in $\theta_0$. For more complex models we use the model-free screening method of \cite{ZhuEtal11} to obtain  $\cS_*$. Under mild conditions on the predictor matrix, Theorem 3 in \cite{ZhuEtal11} ensures the validity of their procedure. In our notations, the result therein states as
\baq\label{eqn:ZhuConsistency}
P( | \cS_* \cap \cS_0^c | \geq r ) \leq \left( 1 - \frac{r}{p+d} \right)^d
\eaq
for positive integers $r$ and $d$: $d = p$ being a good practical choice as per the numerical studies in \cite{ZhuEtal11}.

Combining \eqref{eqn:SISConsistency} or \eqref{eqn:ZhuConsistency} with Corollary~\ref{corollary:BootConsistency} as needed establishes asymptotically accurate recovery of the support set $\cS_0$ through Algorithm~\ref{alg:algoselectboot} in Section~\ref{subsec:Algo}. For classical $n>p$ cases this holds for any statistical model satisfying the conditions (A1) and (A2), whereas in high-dimensional situations it is contingent upon additional conditions required by the screening method used.

\section{Simulation studies}
\label{Section:Simulation}
We now present three simulation studies to examine the empirical performance of our proposed method using model $e$-values, along with a host of other variable selection methods for linear and linear mixed models. In terms of $n$ and $p$, we consider both the classical $n > p$ case (Sections~\ref{subsec:sim1} and \ref{subsec:sim3}) and the high-dimensional case where $p >> n$ (Section~\ref{subsec:sim2}). For all depth calculations, we use halfspace depth \citep{tukey75}.

\subsection{Selecting covariates in linear regression ($n > p$)}
\label{subsec:sim1}
We use the model $Y = X \beta_0 + \epsilon$, with $\epsilon \sim \cN_n(0, I_n)$, $n=1000$ and $p = 60$. We generate the rows of $X$ independently from $\cN_p(0, \Sigma_X)$, with $(\Sigma_X)_{ij} = \rho^{|i-j|}$. In the true coefficient vector $\beta_0$, we set the first 5 coefficients to have non-zero values equal to 1, and the other slope parameters to be zero. We repeat this setup for $\rho \in \{ 0.5, 0.7, 0.9 \}$, and generate 100 independent datasets for each $\rho$. We use the scale-enhanced Bayesian bootstrap of Example \ref{Example:BootExample2} with $\tau_n \in \{ \log n, n^{0.1}, n^{0.2}, n^{0.3}, n^{0.4} \}$, and resampling Monte Carlo size $R = R_{1} = 1000$ for the calculation of bootstrap $e$-values using \eqref{eqn:BootEvalueMC}.

\begin{table}
\centering
\begin{scriptsize}
\begin{tabular}{c|cc|cc|cc}
    \hline
    \multicolumn{7}{c}{Setting 1: $n = 1000, p = 60$}\\\hline
    Choice of $\tau_n$ & \multicolumn{2}{|c}{$\rho = 0.5$} & \multicolumn{2}{|c}{$\rho = 0.7$} & \multicolumn{2}{|c}{$\rho = 0.9$}\\\cline{2-7}
    & Sparsity  & PE $(\times 10^{-4})$ & Sparsity  & PE $(\times 10^{-4})$ & Sparsity  & PE $(\times 10^{-4})$  \\ \hline
    $\tau_n = \log n$       & 5.01  & 4.5   & 5.00  & 3.3   & 5.06  & 2.6    \\
    $\tau_n = n^{0.1}$           & 16.16 & 33.6  & 16.85 & 23.3  & 17.89 & 16.3   \\
    $\tau_n = n^{0.2}$           & 5.74  & 8.1   & 6.03  & 5.7   & 6.75  & 4.8   \\
    $\tau_n = n^{0.3}$           & 5.01  & 4.5   & 5.01  & 3.3   & 4.96  & 5.0   \\
    $\tau_n = n^{0.4}$           & 5.00  & 4.4   & 5.00  & 3.3   & 3.14  & 633.6 \\\hline
    \multicolumn{7}{c}{Setting 2: $n = 60, p = 1000$}\\\hline
    Choice of $\tau_n$ & \multicolumn{2}{|c}{$\rho = 0.5$} & \multicolumn{2}{|c}{$\rho = 0.7$} & \multicolumn{2}{|c}{$\rho = 0.9$}\\\cline{2-7}
    & Sparsity  & PE $(\times 10^{-2})$ & Sparsity  & PE $(\times 10^{-2})$ & Sparsity  & PE $(\times 10^{-2})$  \\ \hline
    $\tau_n = \log n$       	 & 6.63  & 4.5   & 6.34 & 3.4   & 4.79  & 3.7 \\
    $\tau_n = n^{0.1}$      	 & 7.57  & 4.6   & 7.24 & 3.0   & 7.16  & 2.1  \\
    $\tau_n = n^{0.2}$           & 7.38  & 4.6   & 7.23 & 3.0   & 6.61  & 2.3   \\
    $\tau_n = n^{0.3}$           & 6.94  & 4.6   & 6.58 & 3.0   & 5.44  & 3.0   \\
    $\tau_n = n^{0.4}$           & 6.08  & 4.1   & 5.66 & 3.9   & 3.90  & 6.0 \\\hline
    \end{tabular}
\caption{Model sparsity and prediction errors for different choices of $\tau_n$}
\label{table:tabletuning}
\end{scriptsize}
\end{table}

\paragraph{Competing methods}
For comparative evaluation our technique, we consider methods that come from different domains in the model selection literature: (a) The Mixed Integer Optimization (MIO) method of \cite{BertsimasEtal16}, which is based on advanced discrete optimization methods; (b) LASSO and (c) SCAD-penalized linear regression, which represent the domain of penalized regressions; and (d) Stepwise regression using the BIC and forward selection, which is a conventional favorite.

MIO has two versions: all subset and forward selection. The all subset method was much more computationally intensive than other methods, so we used the forward selection MIO. This returns one model for each of the $p+1$ possible sparsity levels of the coefficient vector, following which we select the model with lowest BIC. For the penalized regression methods we use ten-fold cross validation to select the optimal tuning parameter.

\paragraph{Performance measures} We use model sparsity and prediction error on an independently generated dataset $(y_\text{test}, X_\text{test})$ using the same setup as the training data to evaluate each of the methods under consideration. For any estimate $\hat \beta$, these metrics are defined as
$$
\text{Sparsity} ( \hat \beta) = \| \hat \beta \|_0; \quad
\text{PE} (\hat \beta) = \frac{\| X_\text{test} \hat \beta - X_\text{test} \beta_0 \|}{\|X_\text{test} \beta_0 \|}
$$
These are plausible metrics of model evaluation when the features are highly correlated \citep{BertsimasEtal16}.

\paragraph{Choice of tuning parameter}
The top few rows of Table~\ref{table:tabletuning} give the average sparsity and PE values of our $e$-value based method for different values of the bootstrap standard deviation $\tau_n$. Both the metrics take lowest values for $\rho=0.7$ and $\rho = 0.9$ when $\tau_n= \log n$. For $\rho = 0.5$, $\tau_n = n^{0.4}$ has marginally lower values, however it performs much worse than $\tau_n= \log n$ when $\rho = 0.9$.

\begin{figure}
\includegraphics[width=\textwidth]{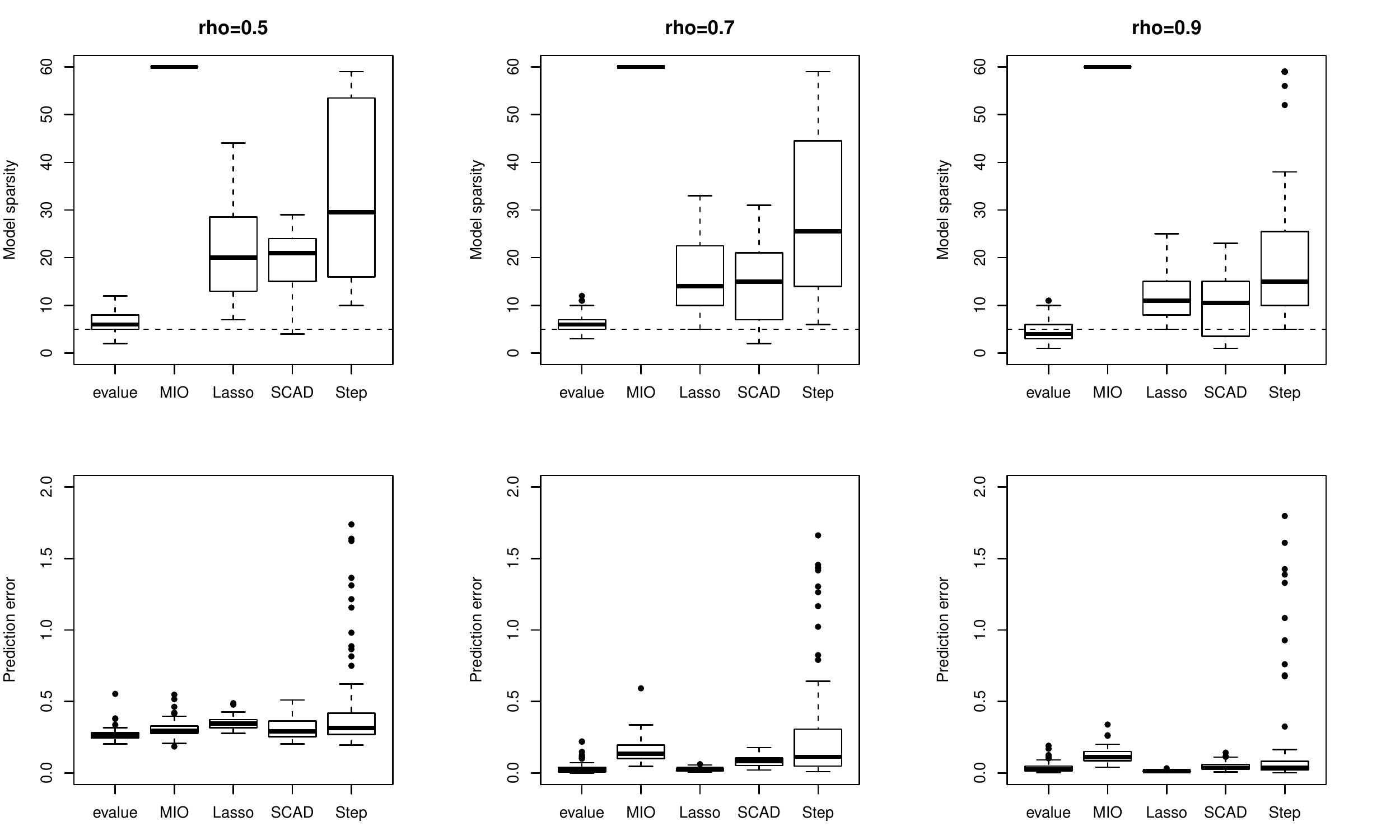}
\caption{Model sparsity (top row) and prediction performance (bottom row) for all methods in $n=60, p=1000$ setup.}
\label{fig:simplotsgamma2}
\end{figure}

\begin{figure}
\includegraphics[width=\textwidth]{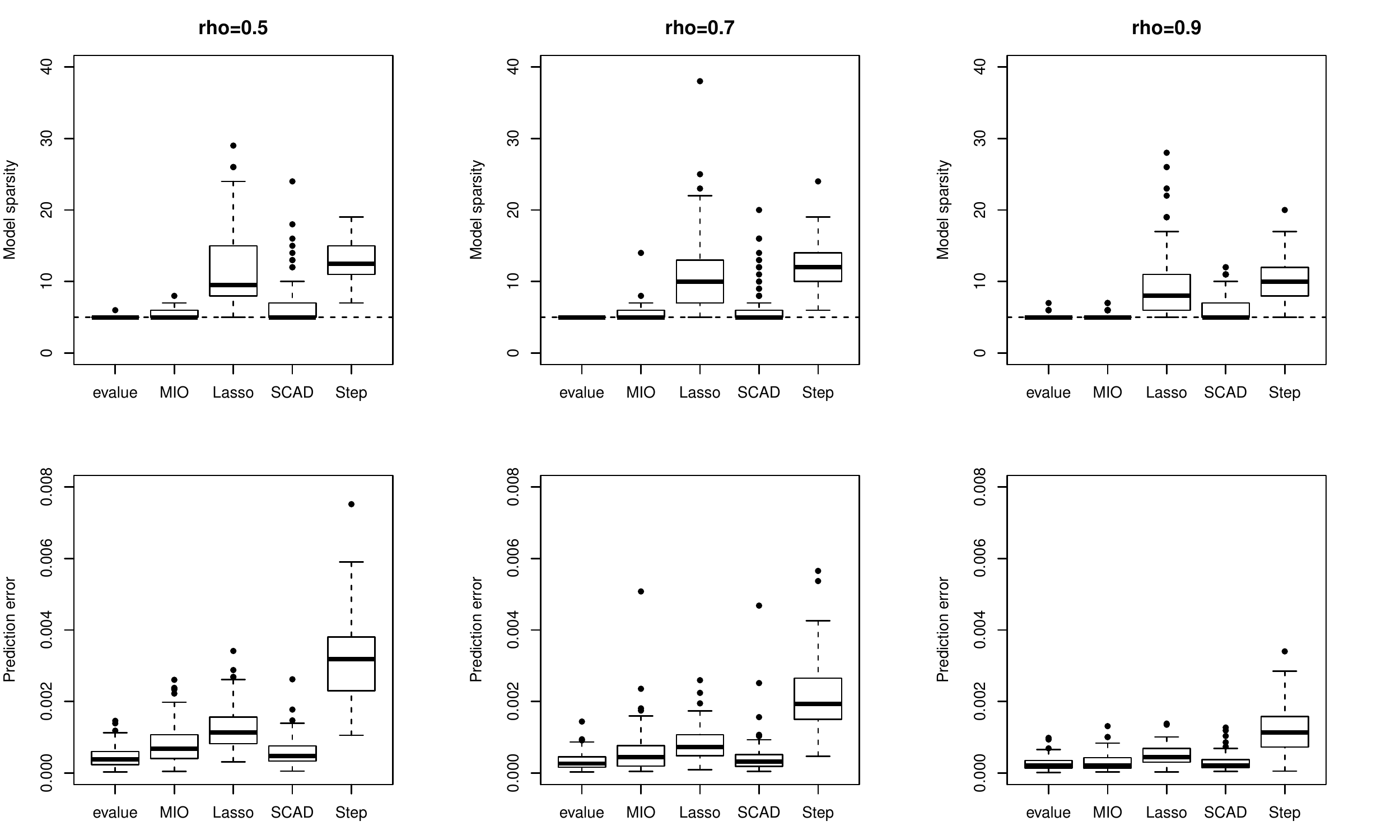}
\caption{Model sparsity (top row) and prediction performance (bottom row) for all methods in $n=1000, p=60$ setup. Dotted line in top row indicates true sparsity level.}
\label{fig:simplotsgamma}
\end{figure}

\paragraph{Comparison with other methods}
Figure~\ref{fig:simplotsgamma} presents boxplots of these performance measures, with $\tau_n = \log n$ assumed for our $e$-value based method. Our method gives the sparsest models and lowest prediction errors among all methods across all values of $\rho$. The second sparsest models are given by MIO, although its prediction errors are higher. On the other hand, prediction errors of SCAD are similar to our method, but our models are more sparse. For all methods, higher values of $\rho$ result in slightly sparser models and higher prediction errors.

\subsection{High-dimensional linear regression ($p >> n$)}
\label{subsec:sim2}
Here we generate the data from the same setup as Section~\ref{subsec:sim1}, but with $n = 60, p = 1000$. 
To implement the $e$-values method, we first perform an initial screening of variables using SIS, then apply our best subset selection procedure.

Among several choices of the tuning parameter, $\tau_n = \log n$ and $\tau_n = n^{0.1}$ are good choices (see second half of Table~\ref{table:tabletuning}). The first choice provides sparser models, while the second choice has lower prediction errors as the correlation among features increases.

To compare with the $n > p$ case, we report the results corresponding to $\tau_n = \log n$ in the boxplots of Figure~\ref{fig:simplotsgamma2}. As compared to the low-dimensional case in Section~\ref{subsec:sim1}, the $e$-values method provides much sparser estimates than other methods. MIO always ends up selecting the least parsimonious model with 60 predictors. Our method has good performance in terms of prediction error as well. Stepwise regression is the most unstable among all five methods across all values of $\rho$ and both performance measures used.

\subsection{Best subset selection in presence of random effects}
\label{subsec:sim3}
Here we use the simulation setup from \cite{ref:PengLu_JMVA12109}. This is a  random intercept-only model:
%
$ \bfY = \bfX \beta + \bfZ \vecU + \bfepsilon$.
%
The data consists of independent groups of observations with multiple observations in each groups,  $\bfZ$ being the within-group random effects design matrix. We consider 9 fixed effects and 4 random effects, with true 
$\bfbeta_0 = (0,1,1,0,0,0,0,0,0)$ and random effect covariance matrix $\Delta$ with elements $(\Delta)_{1 1} = 9$, $(\Delta)_{2 1} = 4.8$, $(\Delta)_{2 2} = 4$, $(\Delta)_{3 1} = 0.6$, $(\Delta)_{3 2} = (\Delta)_{3 3} = 1$, $(\Delta)_{4 j} = 0; j = 1, \ldots, 4$.
%
%
The error variance of $\epsilon$ is set at $\sigma^2 = 1$. The goal is to select covariates of the fixed effect. We use two scenarios for our study: one where the number of subjects ($m$) considered is $30$, and the number of observations in the $i^{\Th}$ subject, $i = 1, \ldots, n$, is $n_i = 5$, and another where $m = 60, n_i = 10$.

Given the original restricted maximum likelihood estimates $\hat\beta, \hat \sigma^2, \hat \Delta$, for the resampling step we use the computational approximation
$$
\hat \beta_r \simeq \hat\beta + \frac{\tau_n}{\sqrt n} 
(\bfX^T \hat \bfV^{-1} \bfX)^{-1} \bfW_r \bfX^T \hat \bfV^{-1} 
(\bfy - \bfX \hat \beta)
$$
with $\bfW_r = \diag (W_{r 1} I_4, \ldots, W_{r 9} I_4)$ and $\hat \bfV = \hat \sigma^2 I_p + \bfZ \hat\Delta \bfZ^T$. This adapts equation \eqref{eqn:BootEqn} in the mixed model setup. We consider $\tau = \tau_{n} / \sqrt n \in \{1, \ldots, 15 \}$ here, and independent Gamma(1,1) random variables as the resampling weights $\{ W_{r1} + 1, \ldots, W_{r9} + 1 \}$.
%
\begin{table}[t]
\centering
\begin{scriptsize}
   \begin{tabular}{ll|lll|lll}
    \hline
    Method      & Tuning     & FPR\% & FNR\% & Model size & FPR\% & FNR\% & Model size \\ \cline{3-8}
    ~ & ~ & \multicolumn{3}{l|}{Setting 1: $n_i=5,m=30$} & \multicolumn{3}{l}{Setting 2: $n_i=10,m=60$}\\ \hline
    $e$-value based  
    ~      & $\tau=3$      & 15.9     & 0.0   & 2.59       & 5.2      & 0.0   & 2.17       \\
    ~      & $\tau=4$      & 8.0      & 0.0   & 2.28       & 2.8      & 0.0   & 2.09       \\
    ~      & $\tau=5$      & 5.2     & 0.0   & 2.18       & 2.0   & 0.0   & 2.06       \\
    ~      & $\tau=6$      & 2.7     & 0.0   & 2.09       & 0.7   & 0.0   & 2.02       \\
    ~      & $\tau=7$      & 2.2   & 0.0   & 2.07       & 0.3   & 0.0   & 2.01       \\
    ~      & $\tau=8$      & 1.5   & 0.0   & 2.05       & 0.3   & 0.0   & 2.01       \\
     \hline
    Peng and Lu \cite{ref:PengLu_JMVA12109}   
    & BIC    & 21.5  & 9.9   & 2.26       & 1.5   & 1.9   & 2.10       \\
    ~      & AIC    & 17    & 11.0  & 2.43       & 1.5   & 3.3   & 2.20       \\
    ~      & GCV    & 20.5  & 6     & 2.30       & 1.5   & 3     & 2.18       \\
    ~      & $\sqrt{\log n/n}$ & 21    & 15.6  & 2.67       & 1.5   & 4.1   & 2.26       \\ \hline
\end{tabular}
\end{scriptsize}
\caption{Comparison between our method and that proposed by \cite{ref:PengLu_JMVA12109} through average false positive percentage, false negative percentage and model size}
\label{table:simtable1}
\end{table}
\begin{table}
\centering
\begin{scriptsize}
    \begin{tabular}{llll}
    \hline
    Method          & ~ & Setting 1 & Setting 2 \\ \hline
    $e$-value based  
    ~               & $\tau=4$ & 79        & 92      \\
    ~               & $\tau=5$ & 87        & 94       \\
    ~               & $\tau=6$ & 93      & 98       \\
    ~               & $\tau=7$ & 94       & 99       \\
    ~               & $\tau=8$ & 96       & 99       \\
    ~               & $\tau=9$ & 97       & 99       \\
    ~               & $\tau=10$ & 98       & 99       \\\hline
    \cite{ref:Bondelletal_Biometrics101069} & ~ & 73        & 83        \\
    \cite{ref:PengLu_JMVA12109}         & ~ & 49        & 86        \\
    \cite{ref:FanLi_AoS122043}           & ~ & 90        & 100       \\ \hline
    \end{tabular}
\end{scriptsize}
\caption{Comparison of our method and three sparsity-based methods of mixed effect model selection through accuracy of selecting correct fixed effects}
\label{table:simtable2MS}
\end{table}

We compare our results with those obtained using the mixed effect model selection method of \cite{ref:PengLu_JMVA12109}, as well as two other methods, all of which are based on sparse estimation of the fixed effect coefficients. In the method proposed by \cite{ref:PengLu_JMVA12109}, the tuning parameter can be selected using several different criteria. We present the false positive percentage (FPR\%), false negative percentage (FNR\%) and model sizes corresponding to four such criteria in Table~\ref{table:simtable1}. Our $e$-value based method handsomely outperforms the  method proposed by \cite{ref:PengLu_JMVA12109}, especially in smaller sample sizes, as long as $\tau \geq 4$. We also compare the percentages of times the correct model was identified, and these results are presented in Table~\ref{table:simtable2MS}, along with the corresponding results from two other papers. The proposed $e$-value based procedure performs best here for $\tau \geq 5$ for the smaller sample setting, and for $\tau\geq 7$ for larger sample setting. 
\section{Eliciting drivers of Indian monsoon precipitation}
\label{Section:IndianMonsoon}
Various  studies indicate that our knowledge about the physical drivers of precipitation in India is incomplete; this is in addition to the known difficulties in modeling precipitation itself \citep{ref:Knuttietal10_JClimate, ref:Trenberth11_ClimateResearch, ref:Trenberth03_BAMS, ref:Wangetal05_GRL}. For example, \cite{goswami} discovered an upward trend in frequency and magnitude of extreme rain events, using daily central Indian rainfall data on a $10^{\circ}$ $\times$ $12^{\circ}$ grid, but  a similar study on a $1^{\circ}\times 1^{\circ}$ gridded data by \cite{ghosh2} suggested  that there are both increasing and decreasing trends of extreme rainfall events, depending on the location. Additionally, \cite{kris} reported increasing trends in exceedances of the $99^{\Th}$ percentile of  daily rainfall; however, there is also a decreasing trend for exceedances of the $90^{\Th}$ percentile data in many parts of India. Significant spatial and temporal variabilities at various scales have also been discovered by \cite{ref:DietzC14} and \cite{ref:DietzC15_ClimateInfo14}.

\begin{figure}
\begin{floatrow}
\capbtabbox{%
\begin{scriptsize}
   \begin{tabular}{l|l}
    \hline
      Predictor dropped     & $e$-value of $\cM_{-j}$       \\ \hline
    - TMAX                  & 0.1490772 \\
    - X120W                 & 0.2190159 \\
    - ELEVATION             & 0.2288938 \\
    - X120E                 & 0.2290021 \\
    - del\_TT\_Deg\_Celsius & 0.2371846 \\
    - X80E                  & 0.2449195 \\
    - LATITUDE              & 0.2468698 \\
    - TNH                   & 0.2538924 \\
    - Nino34                & 0.2541503 \\
    - X10W                  & 0.2558397 \\
    - LONGITUDE             & 0.2563105 \\
    - X100E                 & 0.2565388 \\
    - EAWR                  & 0.2565687 \\
    - X70E                  & 0.2596766 \\
    - v\_wind\_850          & 0.2604214 \\
    - X140E                 & 0.2609039 \\
    - X40W                  & 0.261159  \\
    - SolarFlux             & 0.2624313 \\
    - X160E                 & 0.2626321 \\
    - EPNP                  & 0.2630901 \\
    - TempAnomaly           & 0.2633658 \\
    - u\_wind\_850          & 0.2649837 \\
    - WP                    & 0.2660394 \\ \hline
    none                  & 0.2663496 \\ \hline
    - POL                   & 0.2677756 \\
    - TMIN                  & 0.268231  \\
    - X20E                  & 0.2687891 \\
    - EA                    & 0.2690791 \\
    - u\_wind\_200          & 0.2692731 \\
    - u\_wind\_600          & 0.2695297 \\
    - SCA                   & 0.2700276 \\
    - DMI                   & 0.2700579 \\
    - PNA                   & 0.2715089 \\
    - v\_wind\_200          & 0.2731708 \\
    - v\_wind\_600          & 0.2748239 \\
    - NAO                   & 0.2764488 \\ \hline
    \end{tabular}
\end{scriptsize}
}{%
\caption{Ordered $e$-values when dropping one variable at a time in the Indian summer precipitation data}
\label{table:raintable}
\vspace{5em}
}
\ffigbox{%
  		\includegraphics[width=0.3\textwidth]{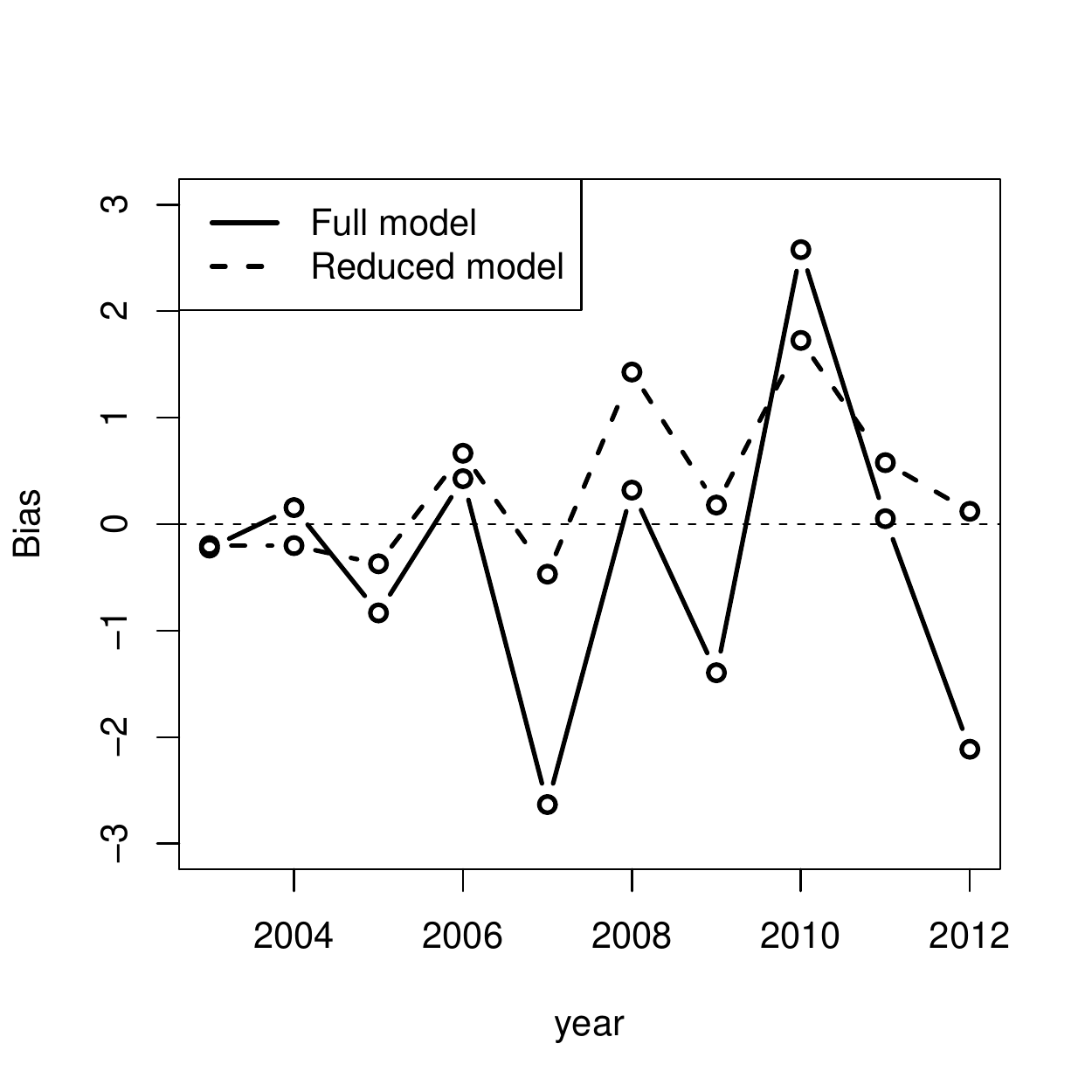}
		\includegraphics[width=0.3\textwidth]{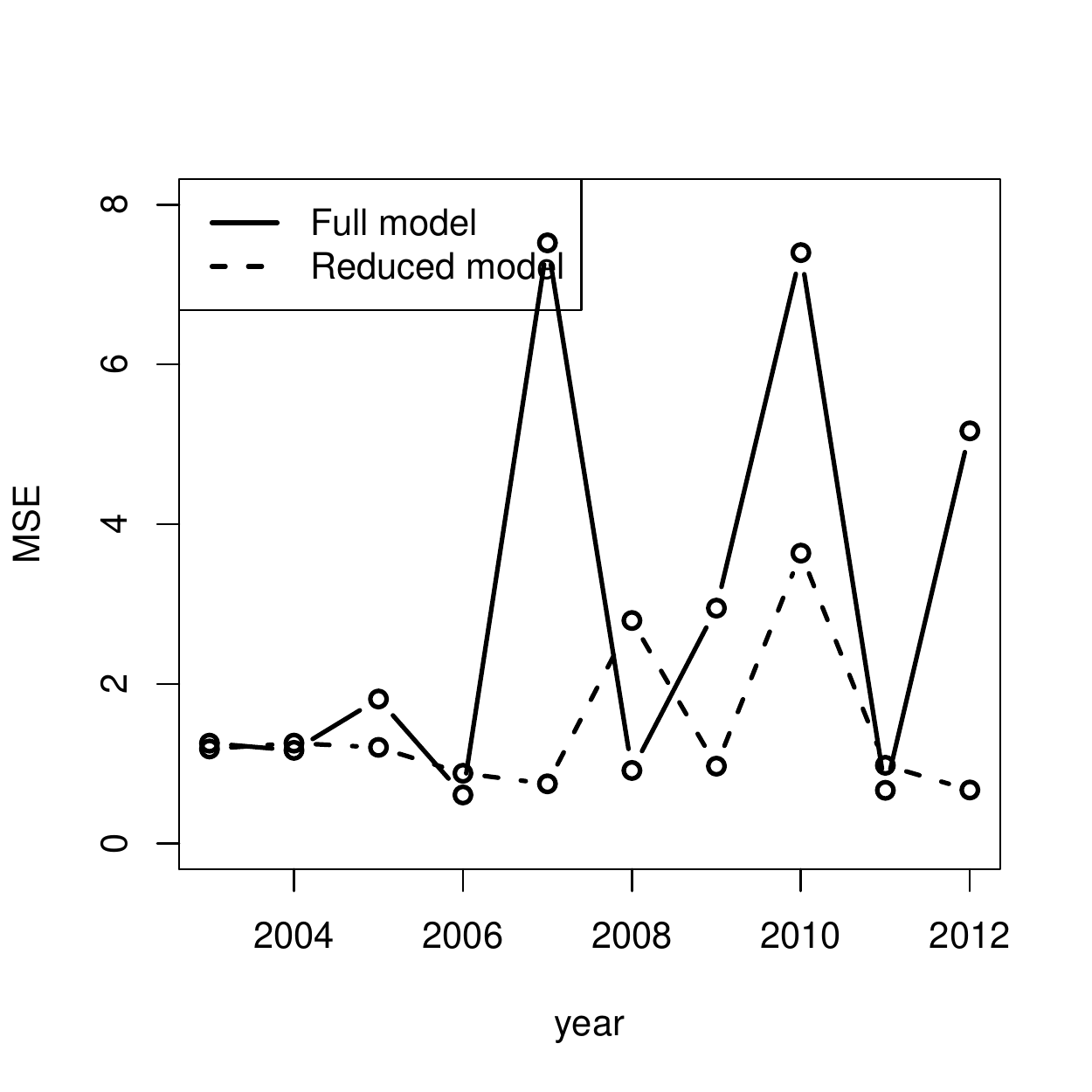}
		\includegraphics[width=0.3\textwidth]{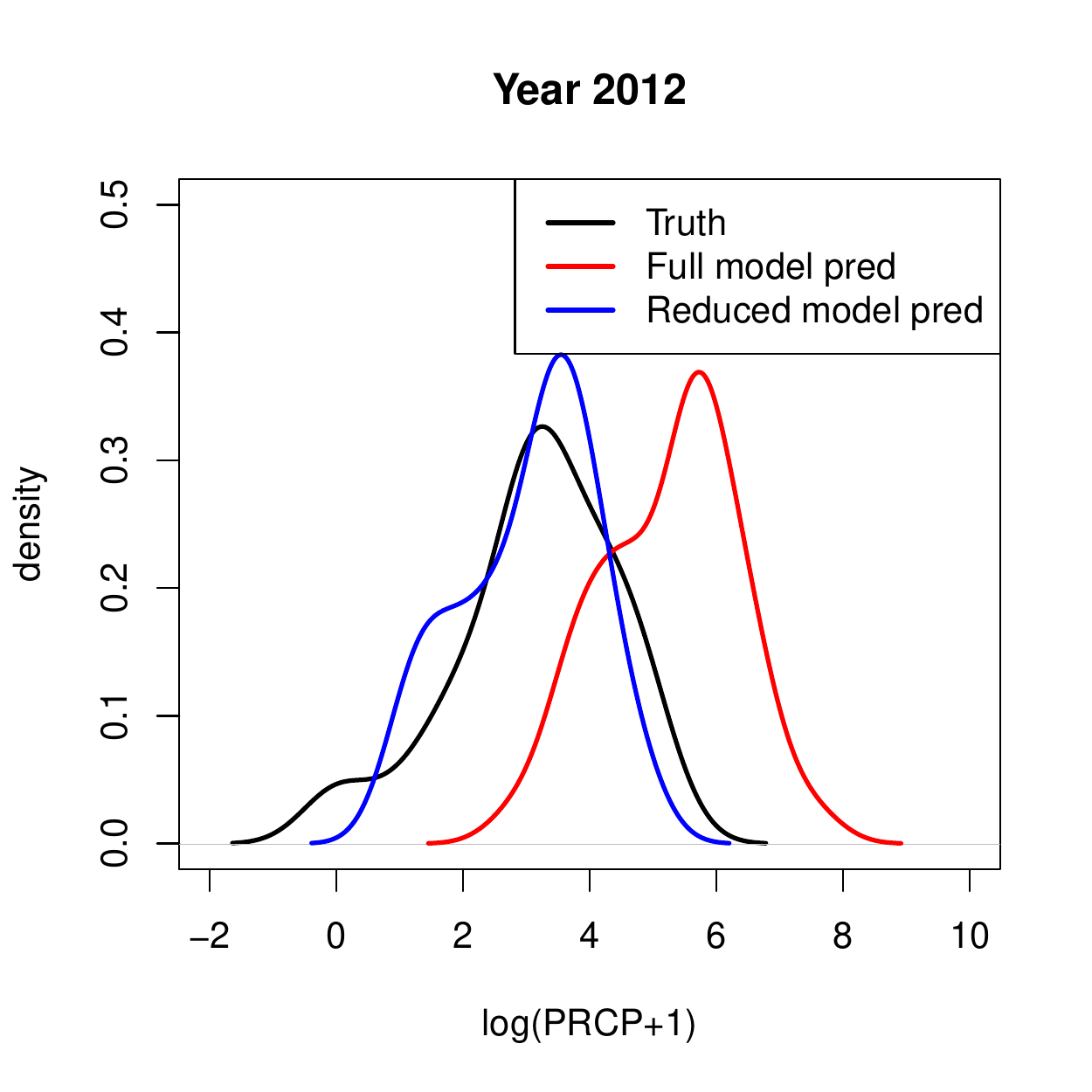}
		\includegraphics[width=0.3\textwidth]{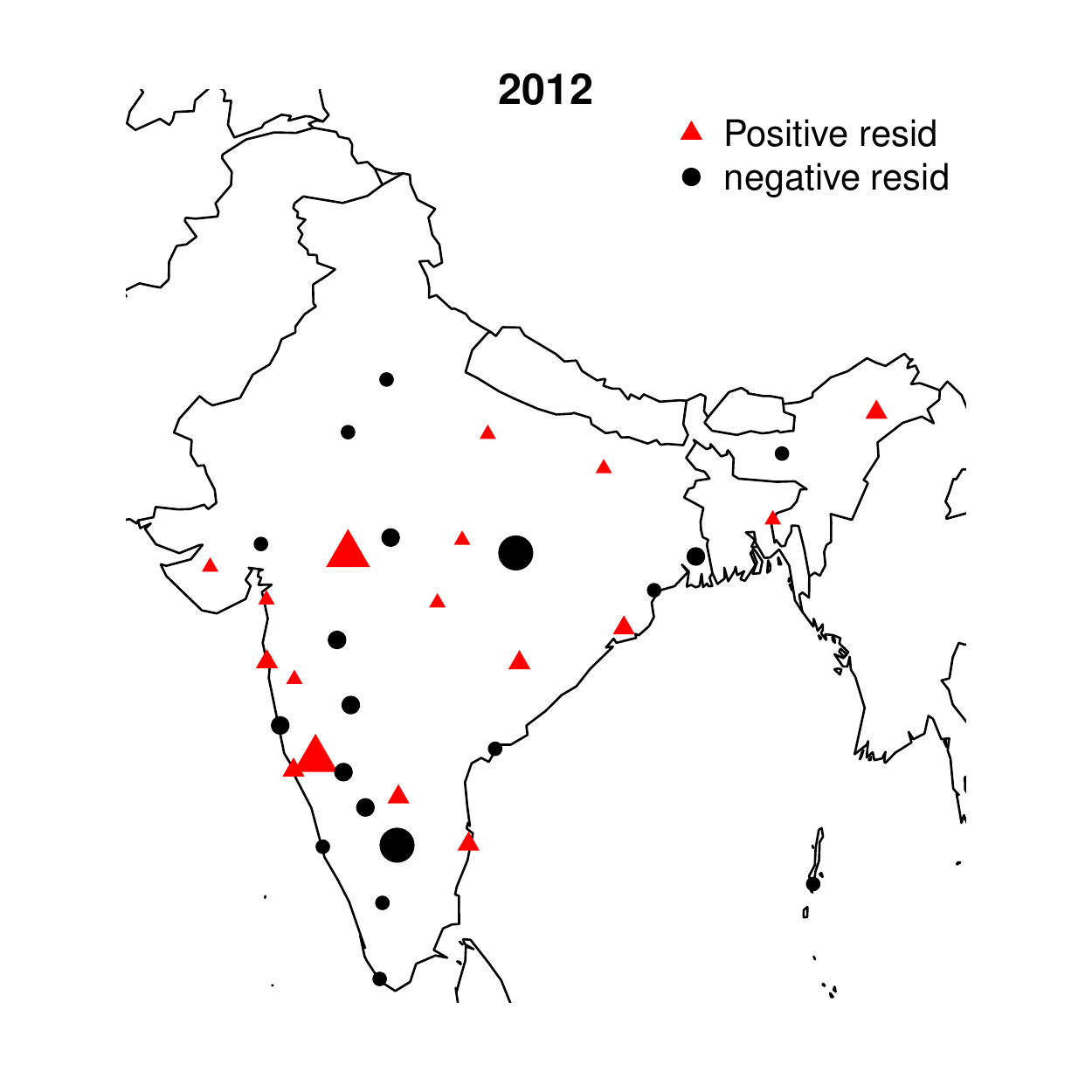}
}{%
\caption{Comparing full model rolling predictions with reduced models: (Top to bottom) Bias across years, Mean Squared Error (MSE) across years, density plots for 2012, stationwise residuals for 2012}
\label{fig:prepost}
}
\end{floatrow}
\end{figure}

Here we attempt to identify the driving factors behind precipitation during the Indian monsoon season using our $e$-value based model selection criterion. Data are obtained from the repositories of the National Climatic Data Center (NCDC) and National Oceanic and Atmospheric Administration (NOAA), for the years 1978--2012. We obtained data on 35 potential covariates of the Indian summer precipitation based on existing knowledge and conjectures from the actual physics driving Indian summer precipitations. Details are provided in the supplementary material. 

We consider annual medians of all covariates as fixed effects, the log yearly rainfall at a weather station as response variable, and include year-specific random intercepts. We use projection depth \citep{zuo03} and resample Monte Carlo sizes $R = R_{1} = 1000$ to calculate bootstrap equivalents of the $e$-values. We train our model on data from the years 1978-2002, run best subset selection for tuning parameter values $\tau_n \in n^k; k \in \{ 0.01, 0.02, \ldots, 0.16 \}$, and select the predictor set corresponding to the $\tau_n$ that minimizes fixed effect prediction errors on the samples from 2003--2012. Table~\ref{table:raintable} lists the estimated $e$-values in increasing order for this full model, as well as all 35 models where a single variable is dropped. Variables listed in the rows above {\textit{none}} in Table~\ref{table:raintable} are considered relevant by our $e$-value criterion. 

All variables selected by our procedure have documented effects on Indian monsoon \citep{KrishChapter, MoonWangHa12}. The single largest contributor is the {\textit{maximum temperature}} variable, whose relation to  precipitation based on the Clausius-Clapeyron relation is now classical knowledge in physics \citep{SingletonToumi12}. It seems that wind velocities high up in the atmosphere are not significant contributors, and the fact that many covariates are selected in the process highlights the complexity of the system. 

%
%
%
%
%
%
%

We checked out-of-sample prediction performance of the estimated model using a rolling validation scheme. For each of the 10 test years: 2003--2012, we select important variables from the model built on past 25 years' data (i.e. use data from 1978--2002 for 2003, 1979-2003 for 2004 and so on), build a model using them and compare predictions on test year obtained from this model with those from the full model. Figure~\ref{fig:prepost} summarizes results obtained through this process. Across all testing years, reduced model predictions have less bias as well as are more stable (top two panels). The better approximations of truth by reduced models is also evident from the density plot for 2012 in the third panel, and the residuals do not show any spatial patterns (bottom panel).

\section{Conclusion}
\label{Section:Conclusion}

We present above an expansive framework and principle, where the definition of a statistical model is broad, and estimation procedures as well as resampling algorithms very general. In such a scenario, we propose a scheme for best subset selection using the newly defined $e$-value. An extremely fast algorithm obtains the true model selection with probability tending to one as the sample size grows by fitting and using only a single model. Simulation results show that our algorithm  performs better than traditional methods in three illustrative examples, and a case study on Indian summer precipitation identifies several important physical drivers of monsoon precipitation. Theoretical consistency results of multiple kinds are provided.

An immediate goal for future research is incorporating \textit{dimension asymptotics} where the parameter dimension is allowed to grow with the sample size, as well as extremely high-dimensional parameters in our framework. Higher order correctness is a possibility in our context since we use resampling methods. The sensitivity of the results to the choice of the depth functions, resampling algorithm and the tuning parameter $\tau_{n}$ associated with it, and the way $D (\hat \theta_{m n}, [\hat \theta_{* n}])$ is summarized to obtain the $e$-value deserve further study. Our results suggest excellent {\textit{asymptotic}} properties that are borne out in the simulations, but finite-sample performance of the procedure needs further scrutiny.

To our knowledge, no other non-penalized method is available in the literature in which computational cost of the covariate selection step scales linearly with dimension of the feature space. This makes the $e$-values a perfect candidate for being an accurate and scalable model selection method in big data. Because our algorithm easily lends itself to parallel computing across different sets of bootstrap samples, it opens up the possibility to be combined with bootstrap methods on subsamples of the full dataset, like the {\it moon}-bootstrap, bag of little bootstraps \citep{KleinerEtal14}, or subsampled double bootstrap \citep{SenguptaEtal16}. For ensemble methods in the machine learning literature, e.g. random forest and bagging, by extension of the same logic $e$-values present an exciting opportunity to explore the formulation of a method for consistent feature ranking.

\section*{Acknowledgements}
This research is partially  supported by the National Science Foundation (NSF) under grants \#IIS-1029711 and 
\#DMS-1622483 and by the National Aeronautics and Space Administration (NASA). The first author also acknowledges the University of Minnesota Interdisciplinary Doctoral Fellowship program.

\appendix 

\section{Proofs of theoretical results}
\label{Section:Proof}

\begin{proof}[Proof of Theorem~\ref{Theorem:ThmRightWrong}]
\textit{Part 1} follows directly from assumption ({\sc B3}).

\textit{Part 2.}
Assuming now that $\cM_n$ is an adequate model, we use ({\sc B1}) property of $D$:
\begin{align}
D ( \hat{\vectheta}_{m n}, [ \hat{\vectheta}_{* n} ]) = 
D \Bigl( \hat{\vectheta}_{m n} - \vectheta_{0 n}, \bigl[ \hat{\vectheta}_{0 n} - \vectheta_{0 n} \bigr] \Bigr),
\end{align}
and decompose the first argument
\begin{align}\label{equation:ThmRightWrongProofPart2Eqn1}
\hat{\vectheta}_{m n} - \vectheta_{0 n} = 
( \hat{\vectheta}_{m n} - \hat{\vectheta}_{0 n} ) 
		+ ( \hat{\vectheta}_{* n} - \vectheta_{0 n} ).
\end{align}
where $\theta_{m n j}$ equals $\theta_{0 n j}$ in indices $j \in \cS_n$ and $C_{n j}$ elsewhere.

Now we have, for any $\cM_n$,
\ban
\hat{\vectheta}_{m n} = \vectheta_{m n} + a_{n}^{-1} T_{m n}
\ean
where $T_{m n}$ is non-degenerate at the $\cS_n$ indices. For the first summand of the right-hand side in \eqref{equation:ThmRightWrongProofPart2Eqn1} we get
\begin{align}\label{equation:ThmRightWrongProofPart2Eqn2}
\hat{\vectheta}_{m n} - \hat{\vectheta}_{* n} = \vectheta_{m n} - \vectheta_{0 n} + R_{n}
\end{align}
where $\BE  \| R_n^2 \| = O ( a_{n}^{-2})$. Since $\cM_n$ is an adequate model, $ \vectheta_{m n} - \vectheta_{0 n} = o(1)$. Thus, substituting the above right-hand side in \eqref{equation:ThmRightWrongProofPart2Eqn1} we get
\begin{align}
\left| D \left( \hat{\vectheta}_{m n} - \vectheta_{0 n}, 
\left[ \hat{\vectheta}_{* n} - \vectheta_{0 n} \right] \right) 
- D \left( \hat{\vectheta}_{* n} - \vectheta_{0 n}, 
\left[ \hat{\vectheta}_{* n} - \vectheta_{0 n} \right] \right) \right|
\leq \| R_{n} \|^{\alpha}
\end{align}
from Lipschitz continuity of $D (\cdot)$ given in (B2). The result now follows. 

{\it Part 3.} 
Since the depth function $E$ is invariant under location and scale transformations, we have
\begin{align}\label{equation:ThmRightWrongProofEqn1}
D ( \hat{\vectheta}_{m n}, [ \hat{\vectheta}_{* n} ]) = 
D \Bigl( a_{n} (\hat{\vectheta}_{m n} - \vectheta_{0 n}), 
\bigl[ a_{n} (\hat{\vectheta}_{* n} - \vectheta_{0 n}) 
\bigr] \Bigr).
\end{align}
Decomposing the first argument,
\begin{align}
a_{n} (\hat{\vectheta}_{m n} - \vectheta_{0 n}) = 
a_{n} (\hat{\vectheta}_{m n} - \vectheta_{m n}) + a_{n} (\vectheta_{m n} - \vectheta_{0 n}). 
\end{align}
Since $\cM_n$ is inadequate, given $\delta > 0$ there exists a subsequence indexed by $ j_n $ such that $| C_{n j_n} - \theta_{0 j_n} | > \delta$ for all $n$. The result follows by application of condition (B4).
\end{proof}

\begin{proof}[Proof of Theorem~\ref{Theorem:ModelScore}]

Several derivations here are similar to those of the proof of Theorem~\ref{Theorem:ThmRightWrong}. Consequently we skip some details.

We have
\ban 
D \Bigl( \hat{\vectheta}_{ r_1 m n}, [ \hat{\vectheta}_{ r * n} ] \Bigr)
& = D \Bigl( \frac{a_n}{\tau_{n}} \bigl( \hat{\vectheta}_{r_1 m n} - \hat{\vectheta}_{* n} \bigr), 
 		\left[ \frac{a_n}{\tau_{n}} \bigl( \hat{\vectheta}_{r * n} - \hat{\vectheta}_{* n} \bigr) \right] \Bigr) \\
& = D \Bigl( 
 \frac{a_n}{\tau_{n}} \bigl( \hat{\vectheta}_{r_1 m n} - \hat{\vectheta}_{ m n} \bigr)
 - \frac{a_n}{\tau_{n}} \bigl( \hat{\vectheta}_{ m n} - \hat{\vectheta}_{* n} \bigr), 
 		\left[ \frac{a_n}{\tau_{n}} \bigl( \hat{\vectheta}_{r * n} - \hat{\vectheta}_{* n} \bigr) \right] \Bigr).
\ean
Conditional on the data, $(a_n / \tau_n) (\hat \theta_{r_1 m n} - \hat \theta_{m n})$ has the same weak limit as $a_n (\hat \theta_{m n} - \theta_{m n})$, and \eqref{equation:ThmRightWrongProofPart2Eqn2} and $\tau_n \rightarrow \infty$ combine to give
$$
\frac{a_n}{\tau_{n}} \bigl( \hat{\vectheta}_{ m n} - \hat{\vectheta}_{* n} \bigr) \stackrel{P}{\rightarrow} 0
$$
as $n \rightarrow \infty$. The results follow directly now.
\end{proof}

\begin{proof}[Proof of Theorem \ref{Theorem:ThmRightNested}]

Since we are dealing with a finite sequence of nested models, it is enough to prove that $e_n(\cM_{1 n} ) > e_n(\cM_{2 n} )$ for large enough $n$, when both $\cM_{1 n}$ and $\cM_{2 n}$ are strictly adequate models and $\cM_{1 n} \prec  \cM_{2 n}$.

Suppose $\BT_0 = \cE (0_p, I_p, g)$. Affine invariance implies invariant to rotational transformations, and since the evaluation functions we consider decrease along any ray from the origin because of (B5), $E ( \vectheta, \BT_0)$ is a monotonocally decreasing function of $ \|\vectheta \|$ for any $\vectheta \in \BR^p$. Now consider the models $\cM^0_{1 n}, \cM^0_{2 n}$ that have 0 in all indices outside $\cS_{1 n}$ and $\cS_{2 n}$, respectively. Take some $\vectheta_{10} \in \matTheta^0_{1 n}$, which is the parameter space corresponding to $\cM^0_{1 n}$, and replace its (zero) entries at indices $j \in \cS_{2 n} \setminus \cS_{1 n}$ by some non-zero $\bfdelta \in \BR^{p - | \cS_{2 n} \setminus \cS_{1 n} |}$. Denote it by $\vectheta_{1 \bfdelta}$. Then we shall have
\begin{align*}
\bftheta_{1 \bfdelta}^T \bftheta_{1 \bfdelta} > \bftheta_{10}^T \bftheta_{10}
& \quad \Rightarrow \quad
D ( \bftheta_{10}, \BT_0) > D ( \bftheta_{1 \bfdelta}, \BT_0)\\
& \quad \Rightarrow \quad
\BE_{s1} D ( \bftheta_{10}, \BT_0) > \BE_{s1} D ( \bftheta_{1 \bfdelta}, \BT_0)
\end{align*}
where $\BE_{s1}$ denotes the expectation taken over the marginal of the distributional argument $\BT_0$ at indices $\cS_{1 n}$. Notice now that by construction $\bftheta_{1 \bfdelta} \in \bfTheta^0_{2 n}$, the parameter space corresponding to $\cM^0_{2 n}$, and since the above holds for all possible $\bfdelta$, we can take expectation 
over indices $\cS_{2 n} \setminus \cS_{1 n}$ in both sides to obtain $\BE_{s1} D( \bftheta_{10},  \BT_0) > \BE_{s2} D ( \bftheta_{20}, \BT_0)$, with $\bftheta_{20}$ denoting a general element in $\bfTheta_{20}$.

Combining (A1) and (A2) we get $a_n \bfV_n^{-1/2} (\hat \bftheta_{* n} - \bftheta_{0 n} ) \leadsto \BT_0$. 
Denote $\BT_n = [ a_n \bfV_n^{-1/2} (\hat \bftheta_{* n} - \bftheta_{0 n} ) ]$, and choose a positive $\epsilon < (\BE_{s1} D ( \bftheta_{10}, \BT_0) - \BE_{s2} D( \bftheta_{20}, \BT_0))/2$. Then, for large enough $n$ we shall have
$$
\left| D ( \bftheta_{10}, \BT_n) - D ( \bftheta_{10}, \BT_0) \right| < \epsilon
\quad \Rightarrow \quad
| \BE_{s1} D ( \bftheta_{10}, \BT_n) - \BE_{s1} D ( \bftheta_{10}, \BT_0) | < \epsilon
$$
following condition (B4). Similarly we have $| \BE_{s2} D( \bftheta_{20}, \BT_n) - \BE_{s2} D( \bftheta_{20}, \BT_0) | < \epsilon $ for the same $n$ for which the above holds. This implies $\BE_{s1} D( \bftheta_{10}, \BT_n) > \BE_{s2} D( \bftheta_{20}, \BT_n)$.

Now apply the affine transformation $\bft (\bftheta_n) = \bfV_n^{1/2} \bftheta_n / a_n + \bftheta_{0 n}$ to both arguments of the depth function above. This will keep the depths constant following affine invariance, i.e. $D(\bft ( \bftheta_{10}), [\hat \bftheta_{* n}]) = D( \bftheta_{10}, \BT_n)$ and $D(\bft ( \bftheta_{20}), [\hat \bftheta_{* n}]) = D( \bftheta_{20}, \BT_n)$. Since this transformation maps $\bfTheta^0_{1 n}$ to $\bfTheta_{1 n}$, the parameter space corresonding to $\cM_{1 n}$, we get $\BE_{s1} D(\bft ( \bftheta_{10}), [\hat \bftheta_{* n}]) > \BE_{s2} D(\bft ( \bftheta_{20}), [\hat \bftheta_{* n}])$, i.e. $e_n ( \cM_{1 n} ) > e_n ( \cM_{2 n} )$.

For any inadequate model $\cM_{j n}, k < j \leq K$, suppose $N_{j n}$ is the integer such that $e_{n_1} ( \cM_{j n_1}) < e_{n_1} (\cM_{* n_1})$ for all $n_1 > N_{j n}$. Part 3 of Theorem~\ref{Theorem:ThmRightWrong} ensures that such an integer exists for every inadequate model. Now define $N = \max_{k < j \leq K} N_{j n}$. Thus $e_{n_1} ( \cM_{* n_1} )$ is larger than $e$-values of all inadequate models $\cM_{j n_1}$ for $k < j \leq K$.
\end{proof}

\begin{proof}[Proof of Corollary~\ref{Corollary:ZeroModelCorollary}]
By construction, $\cM_0$ is nested in all other adequate models in $\BM_0$. Hence Theorem~\ref{Theorem:ThmRightNested} implies $e_n ( \cM_0) > e_n (\cM^{ad}) > e_n (\cM^{inad})$ for any adequate model $\cM^{ad}$ and inadequate model $\cM^{inad}$ in $\mathbb M_0$ and large enough $n$.
\end{proof}

\begin{proof}[Proof of Corollary~\ref{Corollary:AlgoCorollary}]
Consider $j \in \cS_{0}$. Then $\vectheta_{0} \notin \cM_{-j}$, hence $\cM_{-j}$ is inadequate. By choice of $n_1$ from Corollary~\ref{Corollary:ZeroModelCorollary}, $e$-values of all inadequate models are less than 
that of $\cM_{*}$, hence $e_{n_1} (\cM_{-j}) < e_{n_1} (\cM_{*})$.

On the other hand, suppose there exists a $j$ such that $e_{n_1} (\cM_{-j}) \leq e_{n_1} (\cM_*)$ but $j \notin \cS_{0}$. Now $j \notin \cS_{0}$ means that $\cM_{-j}$ is an adequate model. Since $\cM_{-j} $ is nested within $ \cM_{*}$ for any $j$, and the full model is always adequate, we have $e_{n_1} (\cM_{-j}) > e_{n_1} (\cM_{*})$ by 
Theorem~\ref{Theorem:ThmRightNested}: leading to a contradiction and thus completing the proof.
\end{proof}

\begin{proof}[Proof of Corollary~\ref{corollary:BootConsistency}]
Corollary~\ref{Corollary:AlgoCorollary} implies that
$$
\cS_0 = \{j: e_n (\cM_{-j}) < e_n (\cM_*) \}
$$
Now define $ \bar \cS_0 = \{j: e_{r n} (\cM_{-j}) < e_{r n} (\cM_*) \} $. Theorem~\ref{Theorem:ModelScore} implies that
\begin{align*}
e_{r n} (\cM_{-j}) &= e_{n} (\cM_{-j}) + R_{j n}\\
e_{r n} (\cM_{*}) &= e_{n} (\cM_{*}) + R_{* n}\\
\end{align*}
such that $\BE_r | R_{*n} |^2 = o_P(1)$ and $\BE_r | R_{jn} |^2 = o_P(1)$ for all $j$. Hence $P_1( \bar \cS_0 = \cS_0) \rightarrow 1$ as $n \rightarrow \infty$, $P_1$ being probability conditional on the data. Similarly one can prove that the probability conditional on the bootstrap samples that $\bar \cS_0 = \hat \cS_0$ holds goes to 1 as $R, R_1 \rightarrow \infty$, which completes the proof.
\end{proof}


\bibliographystyle{apalike}
\bibliography{DepthModelSelection-AoS}


\end{document}